%% file: m.tex
\documentclass[a4paper,11pt]{article}
\usepackage[utf8]{inputenc}

\usepackage{a4wide}

\usepackage{amsmath,amsfonts,amssymb}

\usepackage{graphicx}
\usepackage{listings}
\usepackage{logic9}
\usepackage{hyperref}
\usepackage{mathtools}
\usepackage{subcaption} 

\usepackage[textsize=small]{todonotes}

\usepackage[electronic]{knowledge}
\knowledgeconfigure{notion, quotation} 
\input{notions}

\usepackage{tikz}
\usepackage{tkz-graph}  
\usetikzlibrary{automata,fit}
\usetikzlibrary{positioning}

\lstdefinelanguage{algo}{%
   morekeywords={function,push,pop,top,for,forall,and,or,not,if,then,else,repeat,until,while,do,report,return,such,that,int,stack,end,delete,let,procedure,call,
   remove, skip, add, choose, forevery}
}

\newcommand{\invariantref}[1]{\hyperref[#1]{#1}}

\title{Revisiting Stateful Partial-Order Reduction}
\author{Frédéric Herbreteau, Sarah Larroze-Jardiné, Gérald Point and Igor Walukiewicz\\
Univ. Bordeaux, CNRS, Bordeaux INP, LaBRI, UMR 5800, 33400, Talence, France}

\input{declarations.tex}

\begin{document}

\maketitle

\input{intro}

\input{related}

\input{systems}

\input{abstract-por}

\input{lex-first}

\input{hard-ifs}

\input{optimal}

\input{pifs}

\input{explore}

\input{experiments}

\input{conclusions}

\bibliographystyle{plain}
\bibliography{m}

\end{document}

%% file: notions.tex
\knowledge{notion}
| transition system
| transition systems
| Transition systems
| \TS

\knowledge{notion}
| reduced transition system
| reduced system
| reduced systems
| reduced transition systems

\knowledge{notion}
| sound
| soundness

\knowledge{notion}
| complete

\knowledge{notion}
| action deterministic

\knowledge{notion}
| acyclic

\knowledge{notion}
| \minTS 

\knowledge{notion}
| remotely potentially includes first

\knowledge{notion}
| \pset

\knowledge{notion}
|terminal state

\knowledge{notion}
| maximal run
| maximal

\knowledge{notion}
| run
| runs

\knowledge{notion}
| full run
| full runs
| full

\knowledge{notion}
| client/server system
| Client/Server systems
| Client/Server system
| client/server systems
| \Prog

\knowledge{notion}
| domain
| \dom
| domains

\knowledge{notion}
| \enabled
| enabled

\knowledge{notion}
| por-equivalence
| por-equivalences
| \bisim
| equivalent
| equivalence
| equivalences

\knowledge{notion}
| Mazurkiewicz trace equivalence
| trace equivalence
| trace equivalent
| equivalence@Mazurkiewicz
| equivalent@Mazurkiewicz
| \sim

\knowledge{notion}
| diamond
| Diamond

\knowledge{notion}
| forward diamond
| Forward diamond

\knowledge{notion}
| independence
| independent

\knowledge{notion}
| dependent
| D
| Da
| De

\knowledge{notion}
| source set
| source sets
| covering source set
| covering source sets
| Source sets

\knowledge{notion}
| \ffirst
| \first
| first
| first set
| first@run

\knowledge{notion}
| first@state
| \First

\knowledge{notion}
| persistent set
| persistent sets
| persistent
| Persistent sets
| \persist

\knowledge{notion}
| trace-optimal
| trace-optimality
| optimal
| optimality
| Trace-optimality
| trace-opimality

 \knowledge{notion}
 | excellent
 | excellent POR algorithm

 \knowledge{notion}
 | \IFS 
 | includes a first-set in

 \knowledge{notion}
 | \PIFS 
 | potentially includes a first set

 \knowledge{notion}
 | \rPIFS 

 \knowledge{notion}
 | I

 \knowledge{notion}
 | -De

 \knowledge{notion}
 | \TS (\Prog )

 \knowledge{notion}
 | lex-run

 \knowledge{notion}
 | lexicographic
 | lex-sequence
 | lex-run
 | lex-runs

 \knowledge{notion}
 | sleep set
 | sleep sets
 | Sleep sets

 \knowledge{notion}
 | sleep-blocked

 \knowledge{notion}
 | non-blocking

 \knowledge{notion}
 | sticking actions
 | sticks

 \knowledge{notion}
 | sleep-invariant

 \knowledge{notion}
 | sticking actions
 | sticks

 \knowledge{notion}
 | \closure 
 | closure

 \knowledge{notion}
| \pclosure 

\knowledge{notion}
| wraps

\knowledge{notion}
| lex-useful

\knowledge{notion}
| subsumption relation
| subsumes
| \subsum 
| subsumption

\knowledge{notion}
| min-closure

\knowledge{notion}
| lex-closure

%% file: declarations.tex
\newcommand{\Prog}{\PP}
\newcommand{\TS}{\mathrm{TS}}
\newcommand{\bisim}{\approx}

\newcommand{\source}{\mathit{source}}

\newcommand{\dar}{\!\downarrow}
\newcommand{\uar}{\!\uparrow}
\newcommand{\be}{\bar{e}}

\newcommand{\bx}{\bar x}
\newcommand{\bth}{\bar\theta}
\newcommand{\bl}{\bar\lambda}
\newcommand{\bC}{\bar C}

\newcommand{\val}{v}
\newcommand{\sact}[1]{\stackrel{#1}{\Longrightarrow}}
\newcommand{\persist}{\mathit{persist}}
\newcommand{\ffirst}{\mathit{first}}
\newcommand{\first}{\mathit{first}}

\newcommand{\dom}{\mathit{dom}}
\newcommand{\Proc}{\mathit{Proc}}
\newcommand{\First}{\mathit{First}}

\newcommand{\minTS}{\mathit{minTS}}
\newcommand{\Alg}{\mathit{Alg}}

\newcommand{\Sleep}{\mathit{sleep}}
\newcommand{\sleep}{\mathit{sleep}}

\newcommand{\Explore}{\mathit{Explore}}
\newcommand{\TreeExplore}{\mathit{TreeExplore}}

\newcommand{\closure}{\text{\emph{closure}}}

\newcommand{\pclosure}{\text{\emph{p-closure}}}
\newcommand{\pset}{\text{\emph{p-set}}}
\newcommand{\enabled}{\text{\emph{enabled}}}

\newcommand{\true}{\text{\emph{true}}}
\newcommand{\false}{\text{\emph{false}}}

\newcommand{\explored}{\text{\emph{Explored}}}

\newcommand{\subsum}{\triangleleft}
\newcommand{\IFS}{\mathit{IFS}}
\newcommand{\PIFS}{\mathit{PIFS}}
\newcommand{\rPIFS}{\mathit{rPIFS}}

\newcommand{\minclosure}{\text{\emph{MinClosure}}}
\newcommand{\lexclosure}{\text{\emph{LexClosure}}}

\newcommand{\chooseaction}{\text{\emph{ChooseAction}}}
\newcommand{\chooseclosure}{\text{\emph{ChooseClosure}}}

\newcommand{\Clients}{\text{\emph{Clients}}}
\newcommand{\Servers}{\text{\emph{Servers}}}

%% file: intro.tex
\textbf{Abstract:}

The goal of partial-order methods is to accelerate the exploration of concurrent systems by examining only a representative subset of all possible runs. The stateful approach builds a transition system with representative runs, while the stateless method simply enumerates them. The stateless approach may be preferable if the transition system is tree-like; otherwise, the stateful method is more effective.

We focus on a stateful method for systems with blocking operations, like locks.
First, we show a simple algorithm with an oracle that is trace-optimal if used as a stateless algorithm.
The algorithm is not practical, though, as the oracle uses an
NP-hard test.
Next, we present a significant negative result showing that in stateful
exploration with blocking, a polynomially close to optimal partial-order
algorithm cannot exist unless P=NP.
This lower bound result justifies looking for heuristics for our simple
algorithm with an oracle.
As the third contribution, we present a practical algorithm going beyond the
standard stubborn/persistent/ample set approach.
We report on the implementation and evaluation of the algorithm.

\section{Introduction}
The goal of partial-order methods is to speed up explicit state exploration of
concurrent systems.
The state space of such systems grows exponentially with
the number of processes.
Fortunately, many runs of a concurrent system can usually be considered equivalent, so it is enough to explore only one run in each equivalence class.
For example, if one process assigns $x:=2$ and another $y:=3$ then the order of execution of these two operations is usually irrelevant; the two sequences are
equivalent, and it is enough to explore only one of the two.
This reduces the number of visited states as well as the time of exploration.
In some cases, the reductions are very substantial.

In recent years, we have seen novel applications of partial-order methods.
One is proving the properties of concurrent
programs~\cite{Far.Commutativity2023}, where equivalence between runs is not only used to reduce the number of proof objectives, but also to simplify proof objectives by choosing particular linearizations.
Similarly, in symbolic executions~\cite{De.Bon.Joh.Pun.Tap.Tve.SymPaths2020},
or testing~\cite{Ozk.Maj.Ora.Trace2019} partial-order can be used to limit the exploration space while still being exhaustive.
Another application is verification of timed systems using recently developed
local-time
zones~\cite{Gov.Her.Sri.Wal.Abstractions2022}.
All these applications, rely on explicit state enumeration, as opposed to
symbolic methods such as SAT or BDDs.

The first partial-order methods were proposed about 35 years ago under the names
of stubborn sets, persistent sets, or ample sets~\cite{Val.Stubborn1991,God.PartialOrder,Kat.Pel.Verification1992}.
The distinctions between the three~\cite{Val.Han.Stubborn2017} are not important for our discussion, so we will focus on persistent sets.
Since this initial works, there has been a continuous stream of research on the
topic. In 2005, Flanagan and Godefroid
introduced \emph{stateless} dynamic reductions~\cite{Fla.God.Dynamic2005},
making stateless methods a focal point of subsequent research.
In 2014, Abdulla et al. proposed a notion of trace-optimality~\cite{Abd.Aro.Jon.Sag.Source2017}, and a race
reversal technique, initiating a new cycle of work on stateless partial-order
methods~\cite{Cha.Cha.Pav.Sin.Vai.Datacentric2018,Aga.Cha.Pat.Pav.Tom.Stateless2021,Jon.Lan.Sag.Awaiting2022,Kok.Mar.Gla.Vaf.Truly2022,Kok.Mar.Vaf.Unblocking2023,Abd.Ati.Das.Jon.Sag.Parsimonious,Kok.Maj.Vaf.Enhancing2024}.

In this paper, we focus on \emph{stateful} partial-order reduction methods,
which have seen relatively less progress over the last two
decades~\cite{Pel.Partialorder2018,Yan.Che.Gop.Kir.Efficient2008,Cot.Pet.Rod.Sou.Quasioptimal2021,Cir.Ene.Far.Mut.Pragmatic2023}.
A partial-order algorithm produces a \emph{reduced transition system} containing a representative for every equivalence class of runs of the system.
A \emph{stateless} approach produces a tree of runs, but saves memory by storing
only one run at a time.
In contrast, a \emph{stateful} approach keeps all visited states in memory so that an exploration can be stopped if a state is revisited.
Each of the two approaches has its advantages.
If the transition system of all the runs is a finite tree, then stateless exploration is preferable.
But when it is not, and this can happen even if the code of each individual
process is a tree, the number of runs may be several orders of magnitude bigger than the number of states, rendering the stateless approach infeasible.
For applications other than reachability testing, as the ones mentioned in the
paragraph above, the stateful approach is usually preferable.

The concept of trace-optimality has significantly influenced developments in stateless model-checking.
An algorithm is \emph{trace-optimal} if it explores exactly one representative
from each equivalence class of runs of the system.
This notion was introduced in~\cite{Abd.Aro.Jon.Sag.Source2017} where it is
simply called optimal.
Recent advances include the design of purely stateless, trace-optimal partial-order algorithms with polynomial memory usage relative to system size~\cite{Kok.Mar.Gla.Vaf.Truly2022,Abd.Ati.Das.Jon.Sag.Parsimonious}.
In comparison, in the stateful approach, a natural goal is to minimize the size of the reduced
transition system.
It is not difficult to see that getting exactly the minimal size is NP-hard.
One of our main results says something much stronger: even approximating the
minimal size within a polynomial factor remains NP-hard.

Blocking is also an important and challenging parameter of modern partial-order
methods.
A system is non-blocking if every permutation of a run preserving the local order of actions of each process remains a valid run.
Systems with only read and write operations generally meet this non-blocking criterion.
However, the introduction of synchronization or locking mechanisms typically invalidates it.
Trace-optimal stateless algorithms are known for non-blocking settings.
The exploration of blocking in the context of stateless
algorithms has only started to be addressed recently~\cite{Kok.Vaf.BAM2021,Jon.Lan.Sag.Awaiting2022,Kok.Mar.Vaf.Unblocking2023}.

The existing stateful partial-order methods identify some persistent sets and
often use them together with sleep sets~\cite{God.PartialOrder} to prevent
redundant trace exploration.
While there are many presentations of persistent sets, there are few places
describing algorithms to actually compute them~\cite{God.PartialOrder,Val.State1998,Far.Klu.Pod.Sound2022}.
Another interesting technique is to use lexicographic ordering on runs~\cite{Kah.Wan.Gup.Monotonic2009}.
The idea is that if we want to explore only one run from an equivalence class, we can as well explore the lexicographically smallest run from each equivalence
class.
Actually, as we observe here, the sleep set mechanism is perfectly suited for doing
precisely this.

In this work, we consider a quite general model of concurrent systems consisting of
servers and clients, both modeled as labeled transition systems.
Every action is a synchronization between a client and a server.
We chose this model because servers can model practically any kind of
communication or synchronization mechanisms like variables, locks, or bounded communication channels.
Importantly, our model does not impose a non-blocking assumption, and all our
methods apply to systems with arbitrary blocking situations.
We differentiate between servers and clients to ensure systems are \emph{acyclic}.
Infinite runs are cumbersome to handle with partial-order techniques, so most recent research on the subject makes the acyclicity assumption.

Our three principal contributions are:
\begin{itemize}
  \item A simple trace-optimal algorithm based on concepts existing in the
  literature: lexicographical ordering and sleep sets. This algorithm isolates a
  concept of ``includes first set'' oracle, $\IFS$, that we consider a conceptual
  contribution of this work (Listing~\ref{lst:lex}).
  \item A strong lower bound showing that assuming $P\not=\NP$ no deterministic algorithm can construct a reduced transition system for a client/server system $\Prog$, of
  polynomial size relative to the
  \emph{minimal size} of a reduced system for $\Prog$, while working in time polynomial in the size of its
  input and its output (Theorem~\ref{thm:por-impossible}).
  \item A new partial-order algorithm based on heuristics to compute
  $\IFS$ oracle (Listing~\ref{lst:full}).
  We show that existing methods in the literature, like computing persistent sets and the race reversal technique, can also be viewed as heuristics for solving the $\IFS$ problem.
\end{itemize}
We have implemented the algorithm and evaluated its performance.

\section{Overview of the paper}
Rather than using a programming language syntax, we opt for transition systems. Each process is modeled as a finite transition system, with
synchronization occurring on shared actions between processes. We distinguish
between two types of processes: clients and servers. Clients must be acyclic,
whereas servers have no such restriction. The acyclicity of clients ensures that all the runs are finite. Each action involves synchronization between one client and one server.

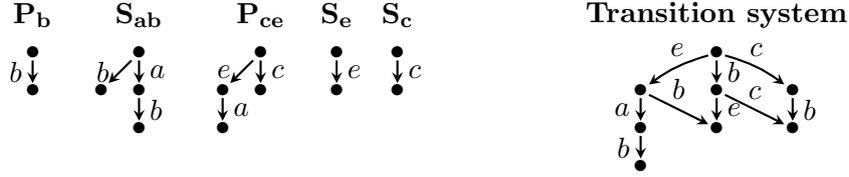
\begin{figure}
  \centering
  \input{figures/fig1.tex}
  \caption{An example of a client/server system (left) with clients $P_b$ and $P_{ce}$, and servers $S_{ab}$, $S_e$ and $S_c$. Processes synchronize on common actions. Its semantics is defined by the transition system to the right.}
  \label{fig:cs-example}
\end{figure}
Figure~\ref{fig:cs-example} presents an example of a client/server system.
We have two clients $P_b$ and $P_{ce}$, and three servers $S_{ab}$, $S_e$, and
$S_c$.
Action $b$ synchronizes $P_b$ and $S_{ab}$, while action $a$ synchronizes
$S_{ab}$ with $P_{ce}$.
We write this as $"\dom"(b)=\set{P_b,S_{ab}}$ and $"\dom"(a)=\set{P_{ce},S_{ab}}$.
The system has runs $eab$, $eb$, $be$, $bc$ and $cb$.

We begin by giving a quite abstract view of partial-order reductions.
This allows us to emphasize the key concepts while keeping the notation relatively light.
The initial concept is a "por-equivalence" relation between runs of a
"client/server system".
This induces the fundamental notion of the "first set" of a "run": $"\first"(u)$ is the set of
first actions of runs equivalent to $u$.
Then a "covering source set" in a state $s$ is simply a set that intersects $\first(u)$ for every "maximal run" $u$ from $s$, with maximal run being one that cannot be extended.
The core idea is that in order to explore all maximal runs from the initial
state modulo por-equivalence, it suffices to follow the actions from covering
source sets (Proposition~\ref{prop:source-covering}).

Coming back to our example.\label{page:overview}
In this paper we will consider Mazurkiewicz "trace equivalence": two adjacent
actions can be permuted if their domains are disjoint.
In our example $bc$ and $cb$ are equivalent as $\dom(b)=\set{P_b,S_{ab}}$ and
$\dom(c)=\set{P_{ec},S_e}$.
We have $"\first"(bc)=\set{b,c}$.
In contrast, the run $eab$ is not equivalent to any other run, so
$"\first"(eab)=\set{e}$.
This shows that $\set{b,e}$ is a covering source set in the initial state of the
system.
In other words, it is not necessary to start an exploration with action $c$ even
though it is enabled in the initial state.

In Section~\ref{sec:lex-first}, we present a very simple trace-optimal
partial-order reduction algorithm (Listing~\ref{lst:lex-tree}).
It uses two known ingredients: "lexicographic" ordering on sequences, and "sleep
sets".
The later are used to stop exploring some actions that are known to have been
explored already. When using sleep sets, it could happen that an exploration
of a run is blocked because all enabled actions are in the sleep-set; we call this
situation a "sleep-blocked" run.
To circumvent such a situation, the algorithm employs an oracle that we call ``includes first set''
oracle.
An oracle test $"\IFS"(s,B)$ asks if there is a maximal run $u$ from $s$ whose first set
is included in $B$, namely $"\first"(u)\incl B$.
The algorithm produces a tree of runs and can be considered stateless.
Subsequently, we modify this algorithm to produce reduced transition systems and not
trees, by using a straightforward "subsumption relation" (Listing~\ref{lst:lex}).

In Section~\ref{sec:ifs-hard}, we show that the $"\IFS"$ test is NP-complete (Proposition~\ref{prop:ifs-is-hard}).
This bears some similarities to the hardness result in~\cite{Cot.Pet.Rod.Sou.Quasioptimal2021}
where it is also shown that some different approach to partial-order reduction
hits a computationally difficult problem.
It turns out though that there is a linear time algorithm for $\IFS$ when
systems do not use synchronization mechanisms (Remark~\ref{rem:non-blocking-case}).
Thus, we have a very simple trace-optimal stateless algorithm in this case.
While this case is often considered in the literature, this restriction is too
strong for us to adopt.

The hardness result for $\IFS$ leaves open whether there is some
other, more clever, partial-order reduction algorithm that is polynomial.
We answer this question negatively in Section~\ref{sec:hard}.
We say that a partial-order algorithm is "excellent" if it works in a time
polynomial in the sum of the sizes of its input and output, producing a reduced
transition system of size only polynomially bigger than the size of the
\emph{minimal} reduced
transition system.
Theorem~\ref{thm:por-impossible} shows that an excellent partial-order algorithm
cannot exist if $P\not=\NP$.
This is the first result of this kind for partial-order methods.

The hardness result justifies our choice of considering the $"\IFS"$ based
algorithm from Listing~\ref{lst:lex} as a reference solution.
In Section~\ref{sec:pifs}, we set out to find some heuristics for $\IFS$.
Our claim is that we have made some progress in Section~\ref{sec:lex-first} even though technically we have proposed an algorithm solving an NP-problem with an NP-oracle.
The progress is that $\IFS$ is a clearly defined problem for which we may have
good intuition on how to approximate it.
We start by giving a characterization of $\IFS$ by the existence of some run
pattern.
The heuristics for $\IFS$ come from a very simple idea:
approximate the existence of a global run by local runs of individual processes.
We propose two heuristic tests, $"\PIFS"$, and $"\rPIFS"$, with $\PIFS$ being
stronger but requiring more pre-computation.

These heuristic tests are used in Section~\ref{sec:algo}, where we propose a new
partial-order reduction algorithm (Listing~\ref{lst:full}).
Our starting point is the idealized algorithm from Listing~\ref{lst:lex}.
We then use $\PIFS$ test instead of $\IFS$.
Since $\PIFS$ test is only a heuristic, we complete it with a method of
computing "covering source sets" inspired by $\rPIFS$ heuristic.
The method does not necessarily compute persistent sets, and it is strictly better than the methods of computing persistent sets we have found in the literature.

In Section~\ref{sec:experiments}, we discuss the results obtained with our
implementation of the algorithm from Listing~\ref{lst:full}.

%% file: figures/fig1.tex
\begin{tikzpicture}[line width=0.8pt,>=stealth]
    \begin{scope}
        \node (Pb) at (0,0.5) {$\mathbf{P_b}$};
        \node[inner sep=0pt] (n0) at (0,0) {$\bullet$};
        \node[inner sep=0pt] (n1) at (0,-0.5) {$\bullet$};
        \draw[->] (n0) edge node[left] {$b$} (n1);
    \end{scope}
    \begin{scope}[xshift=1.4cm]
        \node (Sab) at (0,0.5) {$\mathbf{S_{ab}}$};
        \node[inner sep=0pt] (n0) at (0,0) {$\bullet$};
        \node[inner sep=0pt] (n1) at (-0.5,-0.5) {$\bullet$};
        \node[inner sep=0pt] (n2) at (0,-0.5) {$\bullet$};
        \node[inner sep=0pt] (n3) at (0,-1) {$\bullet$};
        \draw[->] (n0) edge node[left] {$b$} (n1);
        \draw[->] (n0) edge node[right] {$a$} (n2);
        \draw[->] (n2) edge node[right] {$b$} (n3);
    \end{scope}
    \begin{scope}[xshift=3cm]
        \node (Sab) at (0,0.5) {$\mathbf{P_{ce}}$};
        \node[inner sep=0pt] (n0) at (0,0) {$\bullet$};
        \node[inner sep=0pt] (n1) at (-0.5,-0.5) {$\bullet$};
        \node[inner sep=0pt] (n2) at (0,-0.5) {$\bullet$};
        \node[inner sep=0pt] (n3) at (-0.5,-1) {$\bullet$};
        \draw[->] (n0) edge node[left] {$e$} (n1);
        \draw[->] (n0) edge node[right] {$c$} (n2);
        \draw[->] (n1) edge node[right] {$a$} (n3);
    \end{scope}
    \begin{scope}[xshift=4cm]
        \node (Se) at (0,0.5) {$\mathbf{S_e}$};
        \node[inner sep=0pt] (n0) at (0,0) {$\bullet$};
        \node[inner sep=0pt] (n1) at (0,-0.5) {$\bullet$};
        \draw[->] (n0) edge node[right] {$e$} (n1);
    \end{scope}
    \begin{scope}[xshift=4.8cm]
        \node (Se) at (0,0.5) {$\mathbf{S_c}$};
        \node[inner sep=0pt] (n0) at (0,0) {$\bullet$};
        \node[inner sep=0pt] (n1) at (0,-0.5) {$\bullet$};
        \draw[->] (n0) edge node[right] {$c$} (n1);
    \end{scope}
    \begin{scope}[xshift=8cm]
        \node (ss) at (1,0.5){\textbf{Transition system}};
        \node[inner sep=0pt] (n0) at (1,0) {$\bullet$};
        \node[inner sep=0pt] (n1) at (0,-0.5) {$\bullet$};
        \node[inner sep=0pt] (n2) at (1,-0.5) {$\bullet$};
        \node[inner sep=0pt] (n3) at (2,-0.5) {$\bullet$};
        \node[inner sep=0pt] (n4) at (0,-1) {$\bullet$};
        \node[inner sep=0pt] (n5) at (1,-1) {$\bullet$};
        \node[inner sep=0pt] (n6) at (2,-1) {$\bullet$};
        \node[inner sep=0pt] (n7) at (0,-1.5) {$\bullet$};
        \draw[->] (n0) edge[bend right=10] node[above] {$e$} (n1);
        \draw[->] (n0) edge node[right] {$b$} (n2);
        \draw[->] (n0) edge[bend left=10] node[above] {$c$} (n3);
        \draw[->] (n1) edge node[left] {$a$} (n4);
        \draw[->] (n1) edge node[above] {$b$} (n5);
        \draw[->] (n2) edge node[right] {$e$} (n5);
        \draw[->] (n2) edge node[above] {$c$} (n6);
        \draw[->] (n3) edge node[right] {$b$} (n6);
        \draw[->] (n4) edge node[left] {$b$} (n7);
    \end{scope}
\end{tikzpicture}

%% file: related.tex
\section{Related work}\label{sec:related}

The literature on partial-order methods has expanded rapidly in the last decade. In this brief discussion, we focus only on results that are closely related to our work, primarily citing more recent papers.

Partial-order methods have been introduced around
1990-ties~\cite{Val.Stubborn1991,God.Using1991,Kat.Pel.Verification1992}.
The basic principles from these works are still a preferred choice for
applications requiring stateful partial-order
methods~\cite{Far.Klu.Pod.Sound2022}.
Another significant concept for us is the use of lexicographic ordering to identify representative
runs~\cite{Wan.Yan.Kah.Gup.Peephole2008,Kah.Wan.Gup.Monotonic2009,Wac.Kro.Oua.Verifying2013}.
We also highlight a Petri-net unfolding-based
approach~\cite{Rod.Sou.Sha.Kro.Unfoldingbased2015,Cot.Pet.Rod.Sou.Quasioptimal2021} that distinguishes itself by
using prime event structures instead of transition systems.

Stateless partial-order reduction, initiated by~\cite{Fla.God.Dynamic2005},
gained momentum thanks to~\cite{Abd.Aro.Jon.Sag.Source2017}.
This approach has been the focus of extensive research in recent years,
culminating in truly stateless algorithms using only polynomial-sized memory~\cite{Kok.Mar.Gla.Vaf.Truly2022,Abd.Ati.Das.Jon.Sag.Parsimonious}.
Extensions of stateless techniques to systems admitting blocking, like some synchronization mechanisms, have been proposed very
recently~\cite{Jon.Lan.Sag.Awaiting2022,Kok.Mar.Vaf.Unblocking2023}.

The race reversal technique introduced in~\cite{Abd.Aro.Jon.Sag.Source2017} has
been partially adapted for use in stateful
methods~\cite{Yan.Che.Gop.Kir.Efficient2008,Cir.Ene.Far.Mut.Pragmatic2023}.
Additionally, the unfolding based partial-order methods have seen developments like the $k$-partial alternatives~\cite{Cot.Pet.Rod.Sou.Quasioptimal2021}.
On the application front, stateful partial-order methods have been notably used to prove the correctness of concurrent programs~\cite{Far.Klu.Pod.Sound2022,Far.Commutativity2023}, as
well as
for symbolic executions and testing\cite{De.Bon.Joh.Pun.Tap.Tve.SymPaths2020, Ozk.Maj.Ora.Trace2019}.

The true strength of partial-order methods is revealed when using
coarser equivalence relations than Mazurkiewicz traces equivalence~\cite{Mai.Gup.Kan.Maj.Partial2016,Cha.Cha.Pav.Sin.Vai.Datacentric2018,Aga.Cha.Pat.Pav.Tom.Stateless2021,Far.Klu.Pod.Sound2022,Kok.Maj.Vaf.Enhancing2024,Far.Klu.Pod.Stratified2023,Alb.de.Gom.Isa.Stu.Optimal2023}.
Another very fruitful direction has been adapting these methods to weak
memory models ~\cite{Abd.Ati.Jon.Lan.Ngo.Sag.Optimal2019,Kok.Raa.Vaf.Model2019,Kok.Mar.Gla.Vaf.Truly2022}.
We leave these extensions for further work.

Our choice of computational model deviates from recent trends.
Apart from typical programs with variables, partial-order methods have been
adapted to various frameworks, including actor programs~\cite{Tas.Kar.Lau.Leg.Mar.Agh.TransDPOR2012}, event
driven
programs~\cite{Abd.Ati.Bon.Das.Jon.Lan.Sag.Tailoring2023,Jen.Mol.Ray.Dim.Vec.Stateless2015,Mai.Gup.Kan.Maj.Partial2016},
and MPI programs~\cite{Pha.Efficient2019}.
Our choice of a more abstract model aims to simplify the application of our
methods to other models.

%% file: systems.tex
\section{Client/Server systems}

In this section, we formalize our models.
All transition systems in this paper are finite.
They have a unique initial state but no final states.
The edges of a transition system are labeled with actions.
We require action determinism: there is no state with two outgoing transitions labeled with the same action.

\begin{definition}
  A ""transition system"" is a tuple $\struct{S,\Sigma,s^0,\set{\act{a}}_{a\in
        \Sigma}}$, where $S$ is a set of states, $\Sigma$ is the set of actions, $s^0$ is an
  initial state, and $\act{a}$ is the set of transitions between states labeled
  with actions. Transition systems are ""action deterministic"", for every
  $s\in S$ and $a\in \Sigma$ there is at most one $t$ with $s\act{a}t$.

  A transition system is ""acyclic"" if the underlying directed graph (i.e. when edge labels are ignored) is acyclic. A ""terminal state"" is a state without
  outgoing transitions. A ""run"" is a path in the transition system, not
  necessarily from the initial state.
  We write $s\act{v}t$ if there is a run labeled with a sequence of actions $v$
  from $s$ to $t$.
  Sometimes we write just $s\act{v}$ when $t$ is not relevant.
  A ""maximal run"" is a run reaching a terminal state.
  A ""full run"" is a maximal run starting in the initial state.
\end{definition}

A client/server system is a finite collection of finite transition systems synchronizing on common actions.
Each action is a synchronization between one client and one server.
We require that clients are acyclic.

\begin{definition}[Client/Server system]\label{def:cs-system}
  Let $\Proc$ be a finite set of \emph{processes} partitioned into a set of
  \emph{clients} and \emph{servers}: $\Proc=\Clients\cup\Servers$ with $\Clients \cap \Servers = \es$. A
  \emph{""client/server system""}, $"\Prog"$, is a collection of "transition
  systems" $\set{\TS_p}_{p\in\Proc}$, one for each process. Each transition
  system $\TS_p$ has its proper set of actions $\Sigma_p$. Transition systems for
  client processes should be "acyclic". For every action
  $a\in\Sigma=\bigcup_{p\in\Proc} \Sigma_p$ its ""domain"" is a set of processes
  $"\dom"(a)=\set{p : a\in\Sigma_p}$. We require that the "domain" of every action
  in $\Sigma$ consists of one client and one server: $"\dom"(a)=\set{p_c,p_s}$ with
  $p_c\in\Clients$ and $p_s\in\Servers$.
  We write $s_p\act{b}_p$ if there is an outgoing $b$-transition from $s_p$ in $\TS_p$, for
  some process $p$, and action $b$.
  For a sequence of actions $v$ we write $\dom(v)$ for
  $\bigcup\set{\dom(a) : \text{$a$ appears in $v$}}$, and $s_p\act{v}_p$ if
  there is a path labeled $v$ from $s_p$ in $\TS_p$.
\end{definition}

Observe that since every $\TS_p$ in $\Prog$ is action deterministic, and since
every action $a$ appears in exactly one client and one server, then $\TS(\Prog)$
is also action-deterministic.

\begin{definition}
  Semantics of a client/server system $"\Prog"$ is a transition system
  $""\TS(\Prog)""$ whose states are tuples of states of process transitions systems,
  $S=\prod_{p\in\Proc} S_p$; the initial state is the tuple consisting of
  initial states of each process, $s^0=\set{s^0_p}_{p\in\Proc}$; every action
  $a$ synchronizes processes involved in it: $s\act{a} s'$ if $s_p\act{a}_p s'_p$ for
  $p\in\dom(a)$, and $s'_p=s_p$ for $p\not\in\dom(a)$.
  We write $""\enabled""(s)$ for the set of labels of
  transitions outgoing from the  global state $s$ of $"\TS(\Prog)"$.
\end{definition}

We have seen an example of a client/server system $"\Prog"$ in
Figure~\ref{fig:cs-example}.
The transition system $"\TS(\Prog)"$ is shown on the right of the figure.

\begin{remark}
  Because clients are acyclic, and every action involves a client,
  $TS(\Prog)$ is acyclic.
  This is actually the only reason to introduce division into clients and
  servers.
  We need cycles in servers to model read/write operations, but we also want all
  the runs to be finite.
\end{remark}

%% file: abstract-por.tex
\section{Partial-order reduction}\label{sec:por}

We begin by abstractly describing the partial-order method, highlighting the
fundamental concepts as we see them.
Subsequently, we revisit the standard
independence relation from Mazurkiewicz trace theory and instantiate this
abstract formulation.
The remainder of this paper will focus exclusively on
Mazurkiewicz traces. Exploring more coarse equivalence relations
is outside the scope of this work.

The results in this section are not new.
We see them as different formulation of a discussion
from~\cite{Abd.Aro.Jon.Sag.Source2017} where the source sets have been
introduced.

Let us fix a "transition system" $\TS$.
Intuitively it is $"\TS(\Prog)"$ for some client/server system $"\Prog"$, but we do
not need this for our definitions.
An equivalence relation between sequences of actions $u \bisim v$ is called
""por-equivalence"" for $\TS$ if $u\bisim v$ implies that:
\begin{itemize}
  \item $u$ and $v$ have the same length,
  \item if $u$ is a "maximal run" from a state $s$ of $\TS$ then so is $v$,
  \item if $u\bisim v$ then $au\bisim av$ for every action $a$.
\end{itemize}
The requirement on equivalent sequences having the same length simplifies the argument.
Anyway, most equivalences considered in the literature satisfy it.
Relation $\bisim$ may depend on the transition system we examine, but we do
not make it explicit in the definition.

The simplest example of por-equivalence is the identity.
On the other extreme, we can consider relating $u\bisim v$ if they have the same length and
from every state either both $u$ and $v$ are maximal runs ending in the same
state or none of them is a maximal run.
This is a very powerful equivalence if we are interested in reachability, but
it is very difficult to compute too.
Finding good approximations of this equivalence, is an intriguing research
subject.
Another example is "Mazurkiewicz trace equivalence" that we will introduce later
in this section.

The goal of partial-order reduction is to construct for a given client/server
system $"\Prog"$, a reduced transition system
representing all "full runs" in $"\TS(\Prog)"$ (recall that these are "maximal" runs from the initial state).

\begin{definition}\label{def:reducedTS}
  We say that $\TS_r$ is a ""reduced transition system"" for $\TS$ if it is:
  \begin{itemize}
    \item ""sound"": every "full run" of $\TS_r$ is a full run in $\TS$, and
    \item ""complete"": for every full run $u$ in $\TS$ there
          is a full run $v$ in $\TS_r$  with $v\bisim u$.
  \end{itemize}

\end{definition}

Thus, we put an accent on preserving paths up to equivalence as opposed to
just being interested in reachability.
Historically, partial-order reduction has always been interested in verification
of path properties.
Recent applications, such as proving program correctness, also require path preservation.
Moreover, all the existing approaches that are focused on reachability actually
also preserve some kind of por-equivalence.

A general approach to constructing a reduced transition system is to determine
for each state $s$ of $\TS$ a covering source set: a subset of
enabled actions that is sufficient to explore.
For example, if every sequence starting from $b$ is equivalent to a sequence
starting from $c$ then we may choose to include only $b$ in the source
set (cf.~our discussion on page~\pageref{page:overview}).
The notion of the first action modulo $\bisim$ is central for partial-order
reduction. 
\begin{equation*}
  ""\ffirst""(u)=\set{b : \exists v.\ bv\bisim u}
\end{equation*}

We want a source set in a state to be big enough to contain at least one representative from
every equivalence class of "maximal" runs from the state.
Using the above definition this can be formulated as: a source set in $s$ should
intersect every $\ffirst(u)$ for $u$ a "maximal run" from $s$.
We formalize this as follows.

\begin{definition}\label{def:covering-source-set}
  For a "por-equivalence" relation $\bisim$ on paths, and a state $s$ of $\TS$ we define $\First(s)$ as the set of first sets of all maximal runs from $s$:
  \begin{equation*}
    "\First"(s)=\set{"\ffirst"(u): \text{$u$ is a "maximal run" from $s$}}
  \end{equation*}
  A set of actions $B$ is a ""covering source set"" in $s$ if $B\cap
    F\not=\es$ for every $F\in\First(s)$.
\end{definition}
In particular, if  $B$ contains all enabled
actions from $s$ then $B$ is a covering source set at $s$.
Intuitively, smaller covering source sets should give smaller
reduced transition systems.
This is not always true. A bigger but incomparable w.r.t.\ set inclusion
covering source set may give a better reduction.

Observe that the notion of a covering source set depends on the "por-equivalence"
relation $\bisim$, as the definition of $"\ffirst"$ depends on it.

Suppose we have an assignment of a set of actions $\source(s)$ for
every state $s$ of $\TS$.
We can use it to restrict the transition relation to transitions allowed by
source sets: define $s\sact{a} t$ when $s\act{a} t$ and $a\in \source(s)$.
If every $\source(s)$ is a "covering source set" in $s$ then this restricted
transition relation is enough.

\begin{proposition}\label{prop:source-covering}
  Let $\TS$ be a finite acyclic transition system with a transition relation
  $\act{}$.
  Suppose $\sact{}$ is the restricted transition relation derived from a
  "covering source set" assignment.
  For every state $s$ of $\TS$, and every "maximal run" $s\act{u}$ there is
  a run $v\bisim u$ with $s\sact{v}$.
\end{proposition}
\begin{proof}
  The proof is on the length of $u$.
  Suppose $s\act{u}$ is a "maximal run" from $s$.
  Consider $\source(s)$ that is covering by assumption of the lemma.
  This means $"\ffirst"(u)\cap \source(s)\not=\es$.
  Say $b$ is in the intersection.
  By definition of $\ffirst(u)$, there is $bu'\bisim u$.
  By $\bisim$ being "por-equivalence" for $\TS$, $bu'$ is also a maximal run from
  $s$.
  We have $s\sact{b}s'$ by definition of $\sact{}$.
  If $u'$ is empty we are done.
  Otherwise, $s'\act{u'}$ with $u'$ being a maximal run from $s'$.
  Since $bu'$ has the same length as $u$, induction hypothesis applies to $u'$.
  So there is $v'\bisim u'$ with $s'\sact{v'}$.
  This gives $s\sact{b}s'\sact{v'}$ and $bv'\bisim bu'\bisim u$.
  So $bv'\bisim u$ is the sequence required by the proposition.
\end{proof}

This proposition allows us to construct a reduced transition system that is "sound" and "complete", by keeping
only $\sact{}$ transitions and states reachable from the initial state by these
transitions.

\begin{remark}
  We can also consider a more general setting when instead of one $\bisim$
  relation we have a family of relations $\bisim_s$ for each state $s$.
  This corresponds to an idea of dynamic partial-order reduction.
  For example, an assignment $x:=5$ and a test $x>3$ do not commute in
  general,
  but they do in a state where the value of $x$ is $7$.
  Since we do not consider such contextual independence in this work, we
  refrain from presenting this generalization.
\end{remark}

\begin{figure}
  \centering
  \input{figures/fig2.tex}

  \caption{A counterexample to Proposition~\ref{prop:source-covering} in presence of
    infinite runs}
  \label{fig:bad-source}
\end{figure}
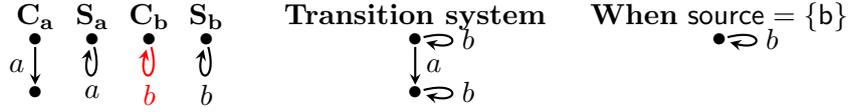

\begin{remark}\label{rem:cycles}
  The above proposition does not hold for transition systems with
  cycles.
  Consider the system to the left of Figure~\ref{fig:bad-source}.
  There are two independent actions $a$ and $b$.
  Client $C_b$ has a cycle on $b$.
  Since every maximal run from the initial state has $b$ on it (see middle picture in Figure~\ref{fig:bad-source}), $\set{b}$ is a
  covering source set in the initial state.
  But this source set does not allow taking a transition labelled $a$. So we
  will never see the run $ab$, as depicted in the picture to the right of Figure~\ref{fig:bad-source}.
  This is why we require that clients are acyclic.
\end{remark}

In this paper we use only the classical Mazurkiewicz "trace
equivalence"~\cite{Maz.Introduction1995}
based on an independence relation induced by domains of actions.

\begin{definition}
  Fix a client/server system $"\Prog"$ with a set of actions $\Sigma$.
  Two actions $a,b\in\Sigma$ are ""independent"" if they have disjoint domains:
  $a ""I"" b$ if $"\dom"(a)\cap"\dom"(b)=\es$.
  Two sequences are ""trace equivalent"", denoted $u"\sim" w$ if $w$ can be
  obtained from $u$ by repeatedly permuting pairs of  adjacent  independent
  actions.
  Two actions are ""dependent"", written $a "D" b$, if they are not independent,
  namely $\dom(a)\cap\dom(b)\not=\es$.
  These notions are extended to sequences: $aI v$ means that $a$ is independent
  of all actions in $v$, and $aD v$ that $a$ is dependent on some action from
  $v$.
  We write $"Da"$ for the set of actions dependent on $a$.
\end{definition}

Directly from definitions we obtain.
\begin{proposition}[\cite{Maz.Introduction1995}]
  For every "client/server system" $\Prog$, "trace equivalence" is a
  "por-equivalence" for $"\TS(\Prog)"$.
\end{proposition}

\subsection{Source sets vs. persistent sets}
The first works on partial-order reduction were built on notions
of stubborn/ample/persistent sets.
For the sake of concreteness let us consider persistent sets
from~\cite{God.PartialOrder}, but the remarks below apply to all three
definitions.

Let us fix some transition system $\TS$.
Recall that persistent set reduction assigns a persistent set of actions to
every state, denoted $\persist(s)$, and explores only actions in the
persistent set in the same way as described above for source sets.
A set of actions $\persist(s)$ is ""persistent"" in $s$ when:
\begin{quote}
  (A2)\quad For every run $s\act{b_1\dots b_n}t$ in TS: if $\forall i.\ b_i\not\in\persist(s)$
  then $\forall i.\ b_i I\persist(s)$.
\end{quote}
Here, $b_i\I\persist(s)$ means that  $b_i$ is "independent" from all
actions in $\persist(s)$.
The property (A2) is the only one we need for our discussion.
Observe that the definition of persistent sets does not give an efficient way of
computing them, since it refers to runs in $\TS$.

\begin{lemma}
  For every state $s$, a "persistent" set in $s$ is a "covering source set" in $s$.
\end{lemma}
\begin{proof}
  Take a "maximal run" $w$ from $s$.
  There must be an action $b\in \persist(s)$ that appears on $w$, as
  otherwise $w "I" \persist(s)$ by (A2), and $wb$ is a "run" by
  "independence".
  This contradicts the fact that $w$ is "maximal".
  So $w=w_1bw_2$ with $w_1\cap \persist(s)=\es$, and $b\in \persist(s)$.
  By (A2) $w_1 "I" b$, so $b\in"\ffirst"(w)$, which means
  $\persist(s)\cap"\ffirst"(w)\not=\es$.
\end{proof}

In the light of this lemma one may ask if the opposite is true: whether every
"covering source set" assignment is a "persistent set" assignment?
This is not the case as the following example shows.

\begin{figure}
  \centering
  \input{figures/fig3.tex}
  \caption{A covering source set may not be a persistent set.}
  \label{fig:not-persistent}
\end{figure}
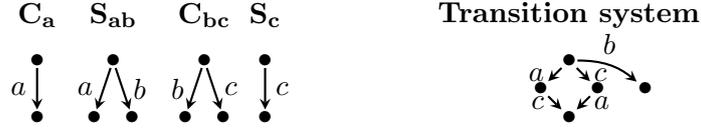

Consider the "client/server system" to the left of Figure~\ref{fig:not-persistent}.
It has two clients ($C_a$ and $C_{bc}$) and two servers ($S_{ab}$ and $S_c$).
Possible runs are $ac$, $ca$, and $b$ as shown by the transition system to the right of Figure~\ref{fig:not-persistent}.
Moreover, $a "I" c$.
Hence, $\set{a,b}$ is a covering  source set in the initial state.
A persistent set containing $\set{a,b}$ needs to include $c$ because of
the run $c$: action $c$ is not in $\set{a,b}$, but it is "dependent" on $b$,  so
by (A2) $c$ should be in the persistent set.

%% file: figures/fig2.tex
\begin{tikzpicture}[line width=0.8pt,>=stealth]
    \begin{scope}
        \node (Ca) at (0,0.3) {$\mathbf{C_a}$};
        \node[inner sep=0pt] (n0) at (0,0) {$\bullet$};
        \node[inner sep=0pt] (n1) at (0,-0.7) {$\bullet$};
        \draw[->] (n0) edge node[left] {$a$} (n1);
    \end{scope}
    \begin{scope}[xshift=0.75cm]
        \node (Sa) at (0,0.3) {$\mathbf{S_a}$};
        \node[inner sep=0pt] (n0) at (0,0) {$\bullet$};
        \draw[->] (n0) edge[loop below] node[below] {$a$} (n0);
    \end{scope}
    \begin{scope}[xshift=1.5cm]
        \node (Cb) at (0,0.3) {$\mathbf{C_b}$};
        \node[inner sep=0pt] (n0) at (0,0) {$\bullet$};
        \draw[->,red] (n0) edge[loop below] node[below] {\textcolor{red}{$b$}} (n0);
    \end{scope}
    \begin{scope}[xshift=2.25cm]
        \node (Sb) at (0,0.3) {$\mathbf{S_b}$};
        \node[inner sep=0pt] (n0) at (0,0) {$\bullet$};
        \draw[->] (n0) edge[loop below] node[below] {$b$} (n0);
    \end{scope}
    \begin{scope}[xshift=5cm]
        \node (ss) at (0,0.3){\textbf{Transition system}};
        \node[inner sep=0pt] (n0) at (0,0) {$\bullet$};
        \node[inner sep=0pt] (n1) at (0,-0.7) {$\bullet$};
        \draw[->] (n0) edge[loop right] node[right] {$b$} (n0);
        \draw[->] (n0) edge node[right] {$a$} (n1);
        \draw[->] (n1) edge[loop right] node[right] {$b$} (n1);
    \end{scope}
    \begin{scope}[xshift=9cm]
        \node (ss) at (0,0.3) {\textbf{When} $\mathsf{source=\set{b}}$};
        \node[inner sep=0pt] (n0) at (0,0) {$\bullet$};
        \draw[->] (n0) edge[loop right] node[right] {$b$} (n0);
    \end{scope}
\end{tikzpicture}

%% file: figures/fig3.tex
\begin{tikzpicture}[line width=0.8pt,>=stealth]
    \begin{scope}
        \node (Ca) at (0,0.6) {$\mathbf{C_a}$};
        \node[inner sep=0pt] (n0) at (0,0) {$\bullet$};
        \node[inner sep=0pt] (n1) at (0,-0.75) {$\bullet$};
        \draw[->] (n0) edge node[left] {$a$} (n1);
    \end{scope}
    \begin{scope}[xshift=1cm]
        \node (Sab) at (0,0.6) {$\mathbf{S_{ab}}$};
        \node[inner sep=0pt] (n0) at (0,0) {$\bullet$};
        \node[inner sep=0pt] (n1) at (-0.25,-0.75) {$\bullet$};
        \node[inner sep=0pt] (n2) at (0.25,-0.75) {$\bullet$};
        \draw[->] (n0) edge node[left] {$a$} (n1);
        \draw[->] (n0) edge node[right] {$b$} (n2);
    \end{scope}
    \begin{scope}[xshift=2.2cm]
        \node (Cbc) at (0,0.6) {$\mathbf{C_{bc}}$};
        \node[inner sep=0pt] (n0) at (0,0) {$\bullet$};
        \node[inner sep=0pt] (n1) at (-0.25,-0.75) {$\bullet$};
        \node[inner sep=0pt] (n2) at (0.25,-0.75) {$\bullet$};
        \draw[->] (n0) edge node[left] {$b$} (n1);
        \draw[->] (n0) edge node[right] {$c$} (n2);
    \end{scope}
    \begin{scope}[xshift=3cm]
        \node (Sc) at (0,0.6) {$\mathbf{S_c}$};
        \node[inner sep=0pt] (n0) at (0,0) {$\bullet$};
        \node[inner sep=0pt] (n1) at (0,-0.75) {$\bullet$};
        \draw[->] (n0) edge node[right] {$c$} (n1);
    \end{scope}
    \begin{scope}[xshift=7cm]
        \node (ss) at (0,0.6) {\textbf{Transition system}};
        \node[inner sep=0pt] (n0) at (0,0) {$\bullet$};
        \node[inner sep=0pt] (n1) at (-0.375,-0.375) {$\bullet$};
        \node[inner sep=0pt] (n2) at (0.375,-0.375) {$\bullet$};
        \node[inner sep=0pt] (n3) at (1,-0.375) {$\bullet$};
        \node[inner sep=0pt] (n4) at (0,-0.75) {$\bullet$};
        \draw[->] (n0) edge node[left] {$a$} (n1);
        \draw[->] (n0) edge node[right] {$c$} (n2);
        \draw[->] (n0) edge[bend left=20] node[above] {$b$} (n3);
        \draw[->] (n1) edge node[left] {$c$} (n4);
        \draw[->] (n2) edge node[right] {$a$} (n4);
    \end{scope}
\end{tikzpicture}

%% file: lex-first.tex
\section{Lex-first exploration and $\IFS$ oracle}\label{sec:lex-first}

As we have seen, the goal of partial-order reduction is to construct a "reduced
transition system" containing at least one "full run" from each equivalence
class of a "por-equivalence" relation. We call a reduced transition system ""trace-optimal"" if no two full runs are equivalent. This notion was proposed in~\cite{Abd.Aro.Jon.Sag.Source2017}, where it was simply called optimal. Since, we aim to construct as small reduced systems as possible, trace-optimality looks like a desired postulate, but it does not guarantee the minimality of the size of a reduced system.

In this section we will present a simple algorithm for constructing
"trace-optimal" systems. It uses three ingredients that we describe later in this
section:
\begin{itemize}
  \item Lexicographic order on sequences allowing to determine a representative
        run for each class of the trace-equivalence relation $"\sim"$.
  \item Sleep sets giving enough information about exploration context.
  \item An oracle ``includes first sets'', denoted $\IFS$, permitting to avoid sleep-blocked
        runs.
\end{itemize}
Lexicographic ordering has been already used in the context of partial-order
reduction~\cite{Kah.Wan.Gup.Monotonic2009}. Sleep sets are one of the classical
concepts for partial-order methods~\cite{God.PartialOrder}.
$\IFS$ oracle is a direct way of avoiding sleep-blocked executions, a well known
challenge in partial-order reduction~\cite{Abd.Aro.Jon.Sag.Source2017}.
Yet, making $\IFS$ explicit will allow us later to develop new heuristics for partial-order reduction.

Let us fix a "client/server system" $"\Prog"$ as in Definition~\ref{def:cs-system}.
Let us also assume that we have some linear ordering on actions of  $\Prog$.
This determines a lexicographic ordering on sequences of actions.
Since we also have our Mazurkiewicz independence relation $"\sim"$ on sequences, and since it
is an equivalence relation, we can use the lexicographic order to define representatives for
equivalence classes of $\sim$ relation.

\begin{definition}
  We say that $w$ is a ""lex-sequence"" if $w$ is the smallest  lexicographically among sequences equivalent to it: that is the smallest sequence in $\set{v : v\sim w}$.
  A ""lex-run"" of $"\TS(\Prog)"$ is a lex-sequence that is a "run" of $"\TS(\Prog)"$.
\end{definition}

We are going to present an algorithm enumerating "full" "lex-runs" of $"\TS"("\Prog")$.
For this we will use the ``includes first set (IFS)'' oracle.
The idea is as follows.
Suppose the algorithm has reached a state $s$ and produced a "sleep set"
$\Sleep(s)$ containing actions that need not be explored from $s$.
We would like to check if there is something left to be explored.
We need to check if the exploration is not ""sleep-blocked"", namely there is a
"maximal run" $u$ from $s$, with
$"\ffirst"(u)\cap\Sleep(s)=\es$.
If there is such run, we need to explore one of $\first(u)$
from $s$.

\begin{definition}
  Let $s$ be a state of $"\TS(\Prog)"$ and $B$ a subset of actions.
  We say that $B$ ""includes a first-set in"" $s$ if there is a "maximal run"
  $u$ from $s$ with  $"\ffirst"(u)\incl B$.
  We write $"\IFS"(s,B)$ when there exists such a maximal run $u$.
\end{definition}

As we will later see, $\IFS(s,B)$ test is \NP-hard.
Eventually we will study approximations of it, but for now we assume it is given
as an oracle.

Listing~\ref{lst:lex-tree} presents a very simple algorithm enumerating all
"full" "lex-runs" of $"\TS(\Prog)"$.
The algorithm gives us also an opportunity to introduce sleep sets.
Each node $n$ of the tree constructed by the algorithm is a pair consisting of a
state of $"\TS(\Prog)"$, denoted $s(n)$, and a set of actions
$\Sleep(n)$.
For readability, we write $\enabled(n)$ instead of $"\enabled"(s(n))$, for the
set of outgoing actions from $s(n)$.

\begin{lstlisting}[caption={Lex exploration with sleep sets, constructs a tree of maximal runs},label=lst:lex-tree,frame=lines,columns=fullflexible,float=ht]
procedure main($"\Prog"$):
  create node $n^0$ with $s(n^0)=s^0$ and $\Sleep(n^0)=\es$
  $\TreeExplore(n^0)$

procedure $\TreeExplore(n)$:  
  $Sl:=\Sleep(n)$      // invariant: $Sl=\Sleep(n)\cup\{$labels of   transitions outgoing from $n\}$
  while $"\enabled"(n)-Sl\not=\es$
      choose smallest $e\in ("\enabled"(n)-Sl)$ w.r.t. linear ordering on actions
      let $s'$ such that $s(n)\act{e} s'$ in $"\TS"("\Prog")$
      if $\IFS(s',"\enabled"(s')-(Sl-De))$£\label{ln:IFS-test}£
        create node $n'$ with $s(n')=s'$ and $\Sleep(n')=Sl-De$
        add edge $n\act{e}n'$
        $\TreeExplore(n')$
      $Sl:=Sl\cup\set{e}$
\end{lstlisting}

""Sleep sets"" are a very elegant mechanism to gather some information about
exploration context.
They can be computed top-down when constructing an exploration graph.
At the root, the sleep set is empty.
For a node $n$ and a transition $n\act{e}n_e$ we have
$\Sleep(n_e)=(\Sleep(n)\cup\set{a_1,\dots,a_k})-De$, where $a_1,\dots,a_k$ are
labels of transitions from $n$ created before the $e$-transition; and
the final $-"De"$ term means that we remove all the actions dependent on $e$.
The intuition behind this formula is as follows.
Assume that we keep an invariant:
\begin{quote}
  (""sleep-invariant""): after exploration of a node $n$ for every "maximal
  run" $u$ from $s(n)$ in $\TS$ such that $\ffirst(u)\cap\Sleep(n)=\es$ we have
  a path $v$
  from $n$ with $u\sim v$.
\end{quote}
Then the formula for $\Sleep(n_e)$ says that we need not explore from $n_e$ a
run starting, say, with $a_1$ if $a_1 "I" e$.
This is because when looking from $n$ these runs have
$a_1$ in their first set, and
they have already been explored from $a_1$ successor of $n$.

The algorithm from Listing~\ref{lst:lex-tree} examines all enabled transitions
in a node $n$ in our fixed order on actions.
For every enabled transition $s(n)\act{e} s'$ it
uses $"\IFS"$ oracle in Line~\ref{ln:IFS-test} to decide if it is neccessary
to explore it.
If the answer is positive, then it creates node $n'$ with an appropriate
"sleep set".
If not, then it skips the transition.
This guarantees that the algorithm is "trace-optimal".

\begin{lemma}\label{lem:lex-optimal-tree}
  The algorithm in Listing~\ref{lst:lex-tree} constructs a tree such that:
  (i) every "full run" in the tree is a full "lex-run" of $"\TS(\Prog)"$, and
  (ii) for every full run $u$ of $\TS(\Prog)$ there is a unique full run $v$ in
  the tree with $v\sim u$.
\end{lemma}
\begin{proof}
  The first step is to show that thanks to "sleep sets" there are no two
  $\sim$-equivalent full runs in the tree.
  Suppose to the contrary that we have $u\sim v$, both full runs in the
  constructed tree.
  Consider the first position where the two sequences differ: $u=wau'$ and
  $v=wbv'$.
  Observe that $a$ and $b$ are "independent", since $u\sim v$.
  Let $n$ be the node reached by the exploration algorithm after reading the
  common prefix $w$.
  We have $n\act{a}n_a$ and $n\act{b}n_b$ in the constructed tree.
  So both $a$ and $b$ are enabled in $n$.
  Assume $a<b$ in our fixed linear ordering on actions. Since we choose the actions according to our
  fixed order, we have $a\in\Sleep(n_b)$.
  As $wau'\sim wbv'$, action $a$ must appear in $v'$; say $v'=v_1av_2$.
  Moreover, $a\in"\first"(v')$ because $au'\sim bv'$.
  Hence, $a "I" v_1$ meaning that the algorithm constructed $n_b\act{v_1}n_1$ with
  $a\in\Sleep(n_1)$, since $a\in\Sleep(n_b)$.
  But then $a$ is "sleep-blocked" and cannot be taken from $n_1$ contrary to the
  assumption that there is a path $wbv_1a$ in the constructed tree.

  To finish the proof we show that every "full" "lex-run" is present in
  the tree.
  For this, it suffices to show that the property "sleep-invariant" above holds.
  We do this by induction on the post-fix DFS order.
  Namely, we take a node $n$ and suppose that the invariant is true for all
  successors of $n$, and all siblings to the left of $n$.
  We show that it holds for $n$.
  So let us take a "maximal" run $u$ from $s(n)$ such that $"\first"(u)\cap\sleep(n)=\es$.
  Consider the smallest $e\in\ffirst(u)$, and the transition $s(n)\act{e}s'$.
  We have $u\sim eu'$ and $\ffirst(u')\incl \ffirst(u)\cup "De"$, where $De$ is the
  set of actions dependent on $e$.
  Let $a_1,\dots,a_k$ be the actions smaller than $e$ for which there are successors of $n$.
  We claim that $e$ successor must be explored, or in other words the test in
  line~\ref{ln:IFS-test} holds.
  Indeed, $\first(u')\cap(\Sleep(n)\cup\set{a_1,\dots,a_k})-De=\es$, so $u'$ is a
  witness that $\IFS(s',\enabled(s')-((\Sleep(n)\cup\set{a_1,\dots,a_k})-De))$ test holds.
  Thus, a transition $n\act{e}n'$ is created.
  By induction assumption, we have a path $v'$ from $n'$ with $v'\sim u'$.
  This gives $ev'\sim eu'$, and we are done.
\end{proof}

We could use this algorithm in stateless model-checking if we had
an implementation of $\IFS(s,B)$.
In a stateful version the algorithm is not very interesting as
trees produced by this algorithm can be, and often are, orders of magnitude bigger
than $"\TS(\Prog)"$.
But we can modify the algorithm to produce a graph instead of a tree.
For this we need to introduce a subsumption relation between nodes.

\begin{definition}
  We say that $n$ ""subsumes"" $n'$, in symbols $n "\subsum" n'$ if the states in
  $n$ and $n'$ are the same, and $\Sleep(n)\incl \Sleep(n')$.
\end{definition}

Observe that if $n$ subsumes $n'$, all runs that are not sleep blocked from $n'$
are also not sleep blocked from $n$. Thus replacing $n'$ by $n$ still yields a "sound" and
"complete" reduced transition system.

\begin{lstlisting}[caption={Lex exploration with sleep sets, constructs a graph},label=lst:lex,frame=lines,columns=fullflexible,float=t]
  procedure main($"\Prog"$):
    create node $n^0$ with $s(n^0)=s^0$ and $\Sleep(n^0)=\es$
    $\explored:=\es$ 
    $\Explore(n^0)$

  procedure $\Explore(n)$:
    $Sl:=\Sleep(n)$ // invariant: $Sl=\Sleep(n)\cup\{$labels of   transitions outgoing from $n\}$
    while $"\enabled"(n)-Sl \not= \es$ £\label{ln:while}£
        choose smallest $e\in ("\enabled"(n)-Sl)$ w.r.t. linear ordering on actions
        let $s'$ such that $s(n)\act{e} s'$ in $"\TS"("\Prog")$
        if $\exists n''\in\explored$ such that $n''"\subsum" (s',Sl - De)$
          add edge $n\act{e} n''$£\label{ln:covering}£
        else if $\IFS(s',"\enabled"(s')-(Sl-De))$
          create node $n'$ with $s(n')=s'$ and $\Sleep(n')=Sl-De$
          add edge $n\act{e}n'$£\label{ln:normal}£
          $\Explore(n')$
        $Sl:=Sl\cup\set{e}$    
    add $n$ to $\explored$  
\end{lstlisting}

In the new algorithm, Listing~\ref{lst:lex}, we test if a node to be
created is subsumed by an already existing node.
If it is, then, instead of creating a new node, we add an edge to the
subsuming node in line~\ref{ln:covering}.
The new algorithm does not satisfy a statement as in
Lemma~\ref{lem:lex-optimal-tree} because edges added due to the subsumption relation may create paths that are not "lex-runs".
We can get a variant of this optimality property when looking at states.
Let us call a state ""lex-useful"" if it appears on a "full" "lex-run".

\begin{lemma}\label{lem:lex-optimal}
  The algorithm in Listing~\ref{lst:lex} constructs a transition system $\TS_r$ such that:
  (i) every "full" "run" in $\TS_r$  is a full "lex-run" of $"\TS(\Prog)"$, and
  (ii) for every full run $u$ of $\TS(\Prog)$ there is a full run $v$ in
  $\TS_r$ with $v\sim u$.
  Moreover, for every node $n$, the state $s(n)$ in this node is "lex-useful".
\end{lemma}
\begin{proof}
  The first statement is direct from the algorithm.

  For the second statement, we verify that the "sleep-invariant" still holds.
  The argument is the same as before with an additional case of edges added in
  line~\ref{ln:covering}.
  For this case we note that if $u$ is a run from $s'$ with $\ffirst(u)\cap
    (Sl-De)=\es$ then by the "sleep-invariant", since $\Sleep(n'')\incl Sl-De$,  there is a
  path from  $n''$ equivalent to $u$.

  For the third statement, observe that every node $n$ in $\TS_r$ is accessible
  by a
  path constructed with edges due to line~\ref{ln:normal}, or in other words
  without edges due to  line~\ref{ln:covering}.
  Such a path is a path in the tree constructed by the algorithm from
  Listing~\ref{lst:lex-tree}.
  So $s(n)$ appears in this tree, implying that it is lex-useful by
  Lemma~\ref{lem:lex-optimal-tree}.
\end{proof}

\begin{remark}\label{rem:use-of-closure}
  In the above algorithms it is not necessary to consider all actions from
  $\enabled(n)$.
  Namely, the ``while'' test in line~\ref{ln:while} can be $\source(n)-Sl\not=\es$ for
  some $\source(n)\incl\enabled(n)$ provided $\source(n)$ is a "covering source
  set" for $s(n)$.
  Indeed, this is all we need to preserve the "sleep-invariant".
  In other words, our algorithm does not need source sets, or persistent sets, but
  can profit from them to reduce the number of $"\IFS"$ tests, or compensate for
  eventual imprecision in an approximation of the $\IFS$ test.

\end{remark}

\begin{remark}\label{rem:sleep-sets-needed}
  Sleep sets  are needed for the optimality result, in a sense
  that some information about exploration context is needed.
  Consider the example presented in Figure~\ref{fig:sleep-needed} (left) that consists in two clients $P_{ac}$ and $P_b$, and two servers $S_{ac}$ and $S_b$.
  \begin{figure}
    \centering
    \input{figures/fig4.tex}
    \caption{Sleep sets are needed for optimality}
    \label{fig:sleep-needed}
  \end{figure}
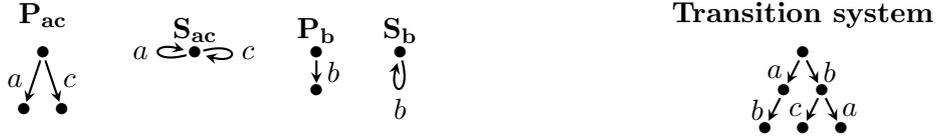
  The runs are depicted on the right of Figure~\ref{fig:sleep-needed}.
  The algorithm first explores the run $ab$, and then in the root state asks if
  there is a run where $a$ is not a first action.
  The answer is positive because of the run $bc$.
  The algorithm then decides to explore $b$ as it is the smallest enabled action.
  At this point it arrives at $n_b$.
  Without a "sleep set" it has no choice but to explore both $a$ and $c$
  from $n_b$.
  But $ba\sim ab$ so this is not "trace-optimal".
  Observe that with sleep sets we get $a\in\Sleep(n_b)$ which is exactly what is
  needed to block unnecessary exploration.
  So our idealized view suggested by Section~\ref{sec:por}
  that we just need to look at a current state and be smart enough to find a good "covering
  source set" is not the whole story.
  On the positive side, the information required about the context is quite
  limited, and easy to compute.
\end{remark}

\begin{remark}\label{rem:ordering-important}
  The ordering on actions can strongly influence the
  size of the result of the above algorithm.
  Consider the system in Figure~\ref{fig:optimal-different-size} where a pair of actions
  with different indexes is "independent".
  \begin{figure}
    \centering
    \input{figures/fig5.tex}
    \caption{A client/server system with two optimal "transition systems", one of linear size and one of exponential size}
    \label{fig:optimal-different-size}
  \end{figure}
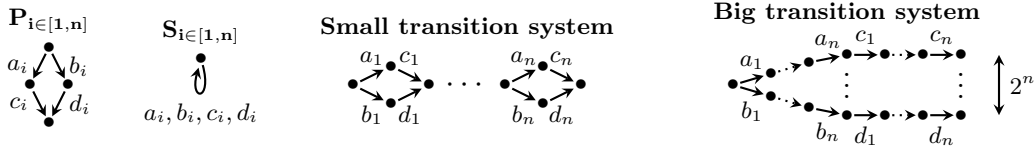
  If we choose to order actions of process $P_i$ before actions of process $P_j$, for
  $i<j$, the algorithm produces a "reduced transition system" of size linear in the number
  of processes.
  If we choose the alphabetic order, $a_i<b_j<c_k<d_l$ for all $i,j,k,l$,
  the algorithm produces a "reduced transition system" of an exponential size.
  Dependence of partial-order methods on the ordering of exploration is a
  well-known issue~\cite{Val.Stop2017}.
\end{remark}

\begin{remark}
  We have used lexicographic order to be able to formulate
  Lemma~\ref{lem:lex-optimal}.
  Actually the algorithm from Listing~\ref{lst:lex-tree} would still be optimal if we choose actions in
  line 9 arbitrarily, and not in some fixed order.
  Without ordering the algorithm from Listing~\ref{lst:lex} would still behave
  like the one from Listing~\ref{lst:lex-tree} without ordering, but the notion of a useful state
  would be less clean.
\end{remark}

\begin{remark}
  The disadvantage of an exploration with "sleep sets" like in
  the algorithm from Listing~\ref{lst:lex} is that some
  states may be duplicated: they may appear several times in the exploration graph with
  different sleep sets.
  Hence, the reduced transition system produced by the algorithm is not
  guaranteed to be smaller than $"\TS(\Prog)"$ which is unsettling.
  It is then interesting to consider a variant of Algorithm~\ref{lst:lex} without sleep
  sets (but still keeping an order on considered successors).
  Unfortunately, Lemma~\ref{lem:lex-optimal} is no longer true in this case.
  We consider the two variants in our experiments (Section~\ref{sec:experiments}).
  Mostly it pays to use sleep sets.
\end{remark}

%% file: figures/fig4.tex
\begin{tikzpicture}[line width=0.8pt,>=stealth]
    \begin{scope}
        \node (Ca) at (0,0.5) {$\mathbf{P_{ac}}$};
        \node[inner sep=0pt] (n0) at (0,0) {$\bullet$};
        \node[inner sep=0pt] (n1) at (-0.25,-0.75) {$\bullet$};
        \node[inner sep=0pt] (n2) at (0.25,-0.75) {$\bullet$};
        \draw[->] (n0) edge node[left] {$a$} (n1);
        \draw[->] (n0) edge node[right] {$c$} (n2);
    \end{scope}
    \begin{scope}[xshift=2cm]
        \node (Sac) at (0,0.3) {$\mathbf{S_{ac}}$};
        \node[inner sep=0pt] (n0) at (0,0) {$\bullet$};
        \draw[->] (n0) edge[loop left] node[left] {$a$} (n0);
        \draw[->] (n0) edge[loop right] node[right] {$c$} (n0);
    \end{scope}
    \begin{scope}[xshift=3.6cm]
        \node (Pb) at (0,0.3) {$\mathbf{P_{b}}$};
        \node[inner sep=0pt] (n0) at (0,0) {$\bullet$};
        \node[inner sep=0pt] (n1) at (0,-0.5) {$\bullet$};
        \draw[->] (n0) edge node[right] {$b$} (n1);
    \end{scope}
    \begin{scope}[xshift=4.7cm]
        \node (Sb) at (0,0.3) {$\mathbf{S_{b}}$};
        \node[inner sep=0pt] (n0) at (0,0) {$\bullet$};
        \draw[->] (n0) edge[loop below] node[below] {$b$} (n0);
    \end{scope}
    \begin{scope}[xshift=10cm]
        \node (ss) at (0,0.5) {\textbf{Transition system}};
        \node[inner sep=0pt] (n0) at (0,0) {$\bullet$};
        \node[inner sep=0pt] (n1) at (-0.25,-0.5) {$\bullet$};
        \node[inner sep=0pt] (n2) at (0.25,-0.5) {$\bullet$};
        \node[inner sep=0pt] (n3) at (-0.5,-1) {$\bullet$};
        \node[inner sep=0pt] (n4) at (0,-1) {$\bullet$};
        \node[inner sep=0pt] (n5) at (0.5,-1) {$\bullet$};
        \draw[->] (n0) edge node[left] {$a$} (n1);
        \draw[->] (n0) edge node[right] {$b$} (n2);
        \draw[->] (n1) edge node[left] {$b$} (n3);
        \draw[->] (n2) edge node[left] {$c$} (n4);
        \draw[->] (n2) edge node[right] {$a$} (n5);
    \end{scope}
\end{tikzpicture}

%% file: figures/fig5.tex
\begin{tikzpicture}[line width=0.8pt,>=stealth,font=\footnotesize]
    \node (Pi) at (0,0.35) {$\mathbf{P_{i \in [1,n]}}$};
    \node[inner sep=0pt] (n0) at (0,0) {$\bullet$};
    \node[inner sep=0pt] (n1) at (-0.25,-0.5) {$\bullet$};
    \node[inner sep=0pt] (n2) at (0.25,-0.5) {$\bullet$};
    \node[inner sep=0pt] (n3) at (0,-1) {$\bullet$};
    \draw[->] (n0) edge node[left] {$a_i$} (n1);
    \draw[->] (n0) edge node[right] {$b_i$} (n2);
    \draw[->] (n1) edge node[left] {$c_i$} (n3);
    \draw[->] (n2) edge node[right] {$d_i$} (n3);
    \begin{scope}[xshift=2cm,yshift=-0.15cm]
        \node (Si) at (0,0.35) {$\mathbf{S_{i \in [1,n]}}$};
        \node[inner sep=0pt] (n0) at (0,0) {$\bullet$};
        \draw[->] (n0) edge[loop below] node[below] {$a_i, b_i, c_i, d_i$} (n0);
    \end{scope}
    \begin{scope}[xshift=4cm,yshift=-0.5cm]
        \node (sg) at (1.5,0.75) {\textbf{Small transition system}};
        \node[inner sep=0pt] (n0) at (0,0) {$\bullet$};
        \node[inner sep=0pt] (n1) at (0.5,0.25) {$\bullet$};
        \node[inner sep=0pt] (n2) at (0.5,-0.25) {$\bullet$};
        \node[inner sep=0pt] (n3) at (1,0) {$\bullet$};
        \draw[->] (n0) edge node[above] {$a_1$} (n1);
        \draw[->] (n0) edge node[below] {$b_1$} (n2);
        \draw[->] (n1) edge node[above] {$c_1$} (n3);
        \draw[->] (n2) edge node[below] {$d_1$} (n3);
        \draw (1.5,0) node {{\large $\cdots$}};
        \node[inner sep=0pt] (n4) at (2,0) {$\bullet$};
        \node[inner sep=0pt] (n5) at (2.5,0.25) {$\bullet$};
        \node[inner sep=0pt] (n6) at (2.5,-0.25) {$\bullet$};
        \node[inner sep=0pt] (n7) at (3,0) {$\bullet$};
        \draw[->] (n4) edge node[above] {$a_n$} (n5);
        \draw[->] (n4) edge node[below] {$b_n$} (n6);
        \draw[->] (n5) edge node[above] {$c_n$} (n7);
        \draw[->] (n6) edge node[below] {$d_n$} (n7);
    \end{scope}
    \begin{scope}[xshift=9cm,yshift=-0.5cm]
        \node (bg) at (1.5,0.95) {\textbf{Big transition system}};
        \node[inner sep=0pt] (n0) at (0,0) {$\bullet$};
        \node[inner sep=0pt] (n1) at (0.5,0.15) {$\bullet$};
        \node[inner sep=0pt] (n2) at (0.5,-0.15) {$\bullet$};
        \node[inner sep=0pt] (n3) at (1,0.3) {$\bullet$};
        \node[inner sep=0pt] (n4) at (1,-0.3) {$\bullet$};
        \node[inner sep=0pt] (n5) at (1.5,0.4) {$\bullet$};
        \node[inner sep=0pt] (n6) at (1.5,-0.4) {$\bullet$};
        \draw[->] (n0) edge node[above] {$a_1$} (n1);
        \draw[->] (n0) edge node[below] {$b_1$} (n2);
        \draw[->,dotted] (n1) edge (n3);
        \draw[->,dotted] (n2) edge (n4);
        \draw[->] (n3) edge node[above] {$a_n$} (n5);
        \draw[->] (n4) edge node[below] {$b_n$} (n6);
        \draw (1.5,0.1) node {{\large $\vdots$}};
        \node[inner sep=0pt] (n7) at (2,0.4) {$\bullet$};
        \node[inner sep=0pt] (n8) at (2,-0.4) {$\bullet$};
        \node[inner sep=0pt] (n9) at (2.5,0.4) {$\bullet$};
        \node[inner sep=0pt] (n10) at (2.5,-0.4) {$\bullet$};
        \node[inner sep=0pt] (n11) at (3,0.4) {$\bullet$};
        \node[inner sep=0pt] (n12) at (3,-0.4) {$\bullet$};
        \draw[->] (n5) edge node[above] {$c_1$} (n7);
        \draw[->] (n6) edge node[below] {$d_1$} (n8);
        \draw[->,dotted] (n7) edge (n9);
        \draw[->,dotted] (n8) edge (n10);
        \draw[->] (n9) edge node[above] {$c_n$} (n11);
        \draw[->] (n10) edge node[below] {$d_n$} (n12);
        \draw (3,0.1) node {{\large $\vdots$}};
        \draw[<->] (3.5,-0.4) -- node[right] {$2^n$} (3.5,0.4);
    \end{scope}
\end{tikzpicture}

%% file: hard-ifs.tex
\section{$\IFS(s,B)$ test is NP-hard}\label{sec:ifs-hard}

In the previous section, we have seen a simple and relatively interesting algorithm for partial-order reduction using $"\IFS"$ oracle.
Unfortunately, there is a serious obstacle to implementing this algorithm because, as we show in this section, the $\IFS(s,B)$ test is NP-hard.
This leads us to study approximations of this test in Section~\ref{sec:pifs}.
In the NP-hardness proof, we construct client/server systems where every process has at most one outgoing transition from every state; this points out that NP-hardness is solely due to blocking.

\begin{proposition}\label{prop:ifs-is-hard} The following problem is \NP-hard:
  given a "client/server system" $"\Prog"$, its global state $s$, and a set of
  actions $B$, does $"\IFS"(s,B)$ hold?
\end{proposition}

\begin{proof}
  Suppose we are given a 3SAT formula $\f$ consisting of $m$ clauses over
  variables $x_1,\dots,x_n$.
  Say the $j$-th clause is of the form $(\a^1_j\lor\a^2_j\lor\a^3_j)$ where
  each $\a$ is either some variable $x_i$ or its negation $\bx_i$.
  We will construct a client/server system $"\Prog"_\f$ such that $\f$ is
  satisfiable if and only if $\IFS(s,B)$ holds for $s$ being the initial state and $B$ some
  set of actions we make precise below.

  All actions in $\Prog_\f$ will implement taking or releasing a lock.
  There is one lock per variable and its negation,
  $\set{x_1,\dots,x_n,\bx_1,\dots,\bx_n}$; one per clause, $\set{c_1,\dots,c_m}$;
  one additional lock for every variable, $\set{e_1,\dots,e_n}$; and a special lock
  $f$.
  For every lock $\ell$ in one of these sets we have an action of taking a
  lock and an action releasing it, denoted respectively $\ell\uar$ and
  $\ell\dar$.

  System $\Prog_\f$ is presented in Figure~\ref{fig:lock-system}.
  For each lock $\ell$ there is a server $S_\ell$ with two states $\set{a,t}$;
  standing for \emph{available} and \emph{taken} respectively.
  Action $\ell\uar$ leads to $t$ and action $\ell\dar$ leads to $a$.
  For every variable $x_i$ there are two client processes $P^0_i$ and $P^1_i$.
  For every clause $(\a^1_j\lor\a^2_j\lor\a^3_j)$ there are client processes
  $C^1_j,C^2_j,C^3_j$.
  Additionally, there are two client processes $D$ and $F$.

  \begin{figure}
    \centering
    \input{figures/fig6.tex}
    \caption{A client/server system $\Prog_\f$ encoding SAT($\f$) for a 3CNF formula $\f$}
    \label{fig:lock-system}
  \end{figure}
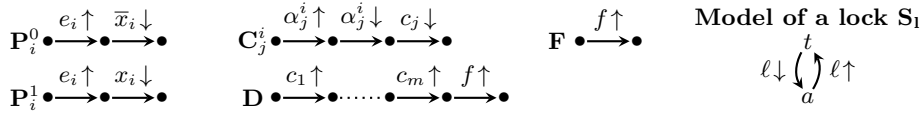

  Consider the state $s$ where every client processes
  is in its  initial state, and where the locks
  $\set{x_1,\dots,x_n,\bx_1,\dots,\bx_n,c_1,\dots,c_m}$
  are taken (they are in state $t$) whereas the other locks are available (they
  are in state
  $a$).
  At this point the only clients enabled are $F$, and  clients corresponding to
  propositional variables, namely $P^0_1,P^1_1,\dots,P^0_n,P^1_n$.
  Client $F$ can take lock $f$.
  For each $i$, either client $P^0_i$ or client $P^1_i$ takes the lock $e_i$ which amounts to choosing a value for variable $x_i$, as either the lock $x_i$ or the lock $\bx_i$ is released, respectively.
  This may allow some client corresponding to a clause to move.
  For example, suppose $x_i$ was released,  and $\a^1_j$, the first literal of
  the $j$-th clause, is $x_i$.
  Then process $C^1_j$ can take $x_i$ and release $x_i$ thus testing if $x_i$
  was available.
  If it was then it can  released $c_j$.
  Releasing $c_j$ intuitively means that the $j$-th clause is satisfied.
  Thus, all locks $c_1,\dots,c_m$ can be released if and only if the formula $\f$ is
  satisfied by the valuation determined by locks from
  $\set{x_1,\dots,x_n,\bx_1,\dots,\bx_n}$ that have been released.
  If all $c_1,\dots,c_m$ are released then client $D$ can take all these locks
  and then take $f$.
  So if $\f$ is satisfiable then there is a maximal run of the system without
  $f\uar$
  in the "first" set.
  If $\f$ is not satisfiable then there is no way to release all the locks
  $c_1,\dots,c_m$, so $D$ cannot take $f$.
  In this case $f\uar$ is in the first set of every maximal run because
  process $F$  can always take $f$.
  This shows that $\IFS(s,\Sigma-\set{f\uar})$ holds iff $\f$ is satisfiable (where $\Sigma$ is the set of all actions in $\Prog_\f$).
\end{proof}

\begin{remark}\label{rem:non-blocking-case}
There is one case where the $"\IFS"$ test can be done in linear time, but our formalism is not convenient to describe it.
We mention it here informally anyway, as it is the case most studied in the context of stateless POR.
In this case, we have threads that can read and write to global variables, but there are no locks or other synchronization primitives.
Moreover, programs are straight-line sequences of instructions.
In this setup, we have the ""non-blocking"" property: if $u$ is a full run, then every permutation of $u$ respecting the thread ordering is also a run.
Clearly, respecting thread ordering is a minimal consistency requirement on permuting actions, as we want that the instructions of each thread execute in
the order specified by the code of the thread.
Observe that this does not mean that every such permutation of $u$ is equivalent to $u$.
Consider, for example, $u=r_1(x),w_2(x,3)$: thread $1$ reading the value of
$x$, and thread $2$ writing $3$ to $x$.
The sequence $w_2(x,3)r_1(x)$ is a permutation of $u$, but it is not
equivalent to $u$, unless the initial value of $x$ is $3$.
In our formalism, we do not have simply $r_1(x)$ action, but need to have
$r_1(x,i)$ actions for all possible values $i$ of $x$.
So the permutation of $r_1(x,0)w_2(x,3)$ becomes $w_2(x,3)r_1(x,3)$; namely, the read action changes as it carries the read value.
This is why formalizing the situation of threads with only reads and writes is rather inconvenient in our setting.
The "non-blocking" property implies that the $\IFS(s,B)$ test can be done by a greedy algorithm: at each step, take the first action of the smallest thread such that the action is either in $B$ or is dependent on some already taken action.
In short, take the action of the smallest thread that is not "sleep-blocked" when started with $\sleep(s)=\enabled(s)-B$.
\end{remark}

%% file: figures/fig6.tex
\begin{tikzpicture}[line width=0.8pt,>=stealth,font=\footnotesize]
    \begin{scope}
        \begin{scope}
            \node (Pi0) at (-0.3,0) {$\mathbf{P}_i^0$};
            \node[inner sep=0pt] (n0) at (0,0) {$\bullet$};
            \node[inner sep=0pt] (n1) at (0.75,0) {$\bullet$};
            \node[inner sep=0pt] (n2) at (1.5,0) {$\bullet$};
            \draw[->] (n0) edge node[above] {$e_i\uar$} (n1);
            \draw[->] (n1) edge node[above] {$\bx_i\dar$} (n2);
        \end{scope}
        \begin{scope}[yshift=-0.75cm]
            \node (Pi1) at (-0.3,0) {$\mathbf{P}_i^1$};
            \node[inner sep=0pt] (n0) at (0,0) {$\bullet$};
            \node[inner sep=0pt] (n1) at (0.75,0) {$\bullet$};
            \node[inner sep=0pt] (n2) at (1.5,0) {$\bullet$};
            \draw[->] (n0) edge node[above] {$e_i\uar$} (n1);
            \draw[->] (n1) edge node[above] {$x_i\dar$} (n2);
        \end{scope}
    \end{scope}
    \begin{scope}[xshift=3cm]
        \begin{scope}
            \node (Cij) at (-0.3,0) {$\mathbf{C}^i_j$};
            \node[inner sep=0pt] (n0) at (0,0) {$\bullet$};
            \node[inner sep=0pt] (n1) at (0.75,0) {$\bullet$};
            \node[inner sep=0pt] (n2) at (1.5,0) {$\bullet$};
            \node[inner sep=0pt] (n3) at (2.25,0) {$\bullet$};
            \draw[->] (n0) edge node[above] {$\a^i_j\uar$} (n1);
            \draw[->] (n1) edge node[above] {$\a^i_j\dar$} (n2);
            \draw[->] (n2) edge node[above] {$c_j\dar$} (n3);
        \end{scope}
        \begin{scope}[yshift=-0.75cm]
            \node (D) at (-0.3,0) {$\mathbf{D}$};
            \node[inner sep=0pt] (n0) at (0,0) {$\bullet$};
            \node[inner sep=0pt] (n1) at (0.75,0) {$\bullet$};
            \node[inner sep=0pt] (n2) at (1.5,0) {$\bullet$};
            \node[inner sep=0pt] (n3) at (2.25,0) {$\bullet$};
            \node[inner sep=0pt] (n4) at (3,0) {$\bullet$};
            \draw[->] (n0) edge node[above] {$c_1\uar$} (n1);
            \draw[dotted] (n1) -- (n2);
            \draw[->] (n2) edge node[above] {$c_m\uar$} (n3);
            \draw[->] (n3) edge node[above] {$f\uar$} (n4);
        \end{scope}
    \end{scope}
    \begin{scope}[xshift=7cm]
        \node (F) at (-0.3,0) {$\mathbf{F}$};
        \node[inner sep=0pt] (n0) at (0,0) {$\bullet$};
        \node[inner sep=0pt] (n1) at (0.75,0) {$\bullet$};
        \draw[->] (n0) edge node[above] {$f\uar$} (n1);
    \end{scope}
    \begin{scope}[xshift=10cm]
        \node (Sl) at (0,0.3) {\textbf{Model of a lock $\mathbf{S_l}$}};
        \node[inner sep=1pt] (n0) at (0,0) {$t$};
        \node[inner sep=1pt] (n1) at (0,-0.75) {$a$};
        \draw[->] (n0) edge[bend right=30] node[left] {$\ell\dar$} (n1);
        \draw[->] (n1) edge[bend right=30] node[right] {$\ell\uar$} (n0);
    \end{scope}
\end{tikzpicture}

%% file: optimal.tex
\section{Partial-order reduction is $\NP$-hard}\label{sec:hard}

Trace-optimality discussed in Section~\ref{sec:lex-first} is an interesting concept,
and even a guiding principle for stateless algorithms, but it is just an
approximation of what we want if we are interested in constructing "reduced
transition systems".
Indeed, the example from Figure~\ref{fig:optimal-different-size} shows two
trace-optimal transition systems, one of a linear size and the other of an exponential size.

We have seen that $"\IFS"$ test is NP-hard, so we cannot hope to implement
the algorithm from Listing~\ref{lst:lex}.
Yet this lower-bound on $\IFS$ does not imply that there does not exist some
other method that is guaranteed to construct small "reduced transition systems".
In this section, we formalize what this can mean and show that this is impossible assuming $P\not=\NP$.

Recall that the goal of partial-order reduction is to construct for a given
system $"\Prog"$, a "reduced system" that is "sound" and "complete". Among such systems, we would ideally like to construct one with the smallest number of states. We write $""\minTS""(\Prog)$ for the smallest number of states of a
"sound" and "complete" "reduced transition system" for $"\Prog"$. There may be several non-isomorphic "transition systems" with a minimal number of states.

In this section, we show that there is no polynomial-time algorithm for constructing a reduced transition system that is polynomially close to optimal, even when the size of the output is taken into account when counting the complexity.

\begin{definition}\label{def:excellent}
	We say that $\Alg$ is an ""excellent POR algorithm"" if there are polynomials
	$q(x)$ and $r(x)$ such that given a "client/server system" $"\Prog"$, the algorithm constructs a
	"sound" and "complete" "transition system" for $\Prog$ of size bounded by $q(|"\minTS"(\Prog)|)$ in
	time $r(|\Prog|+|"\minTS"(\Prog)|)$.
\end{definition}

The main result is:

\begin{theorem}\label{thm:por-impossible}
	If P$\not=$NP then there is no "excellent POR algorithm".
\end{theorem}
The rest of this section is devoted to the proof of this theorem.

We provide a reduction from $3$-SAT problem.
Consider a formula over literals $\set{x_1,\dots,x_n}\cup\set{\bx_1,\dots,\bx_n}$ and with $k$ clauses:
\begin{equation*}
	\f=(\a^1_1\lor\a^1_2\lor\a^1_3)\land\dots \land(\a^k_1\lor\a^k_2\lor\a^k_3)
\end{equation*}
where each $\a^i_j$ is a literal.

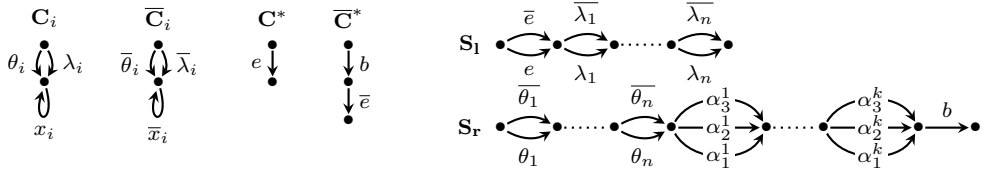
\begin{figure}
	\centering
	\input{figures/fig7.tex}
	\caption{Program $\Prog_\f$ for the lower bound argument.}
	\label{fig:prog-phi}
\end{figure}

Client/server system $\Prog_\f$ is presented in Figure~\ref{fig:prog-phi}.
It has several clients and two servers $S_l$ and $S_r$.
For every $i=1,\dots,n$ we have a client $C_i$ corresponding to variable $x_i$.
Client $C_i$ has two states: top and bottom.
There are two transitions from top to bottom state, the $\theta_i$ transition
intuitively says that $x_i$ should be true, and $\lambda_i$ that it should be false.
This choice is not encoded in the reached state though, as the two transitions
go to the same bottom state where a transition on $x_i$ is possible.
Similarly, for client $\bC_i$ but now we have $\bth_i$, $\bl_i$, and $\bx_i$
actions.
There are also two special clients $C^*$ and $\bC^*$.
The first does $e$ that is enabled in the initial state.
The second can do $\be$, but only  after doing $b$, and this in turn can
happen only when server $S_r$ terminates.

Among the two servers, only $S_r$ depends on formula $\f$.
Server $S_l$ synchronizes on $\l_i$ actions, and $S_r$ on $\th_i$ actions.
Server $S_l$ starts with a choice between $e$ and $\be$, while the second part
of $S_r$ corresponds to the clauses of the formula $\f$, and finishes with $b$ action.

\begin{lemma}\label{lem:only-e}
	If $\f$ is not SAT then all runs of $\Prog_\f$ start with $e$. In other words,
	$e\in\ffirst(w)$ for every "full" run $w$ of $\Prog_\f$.
\end{lemma}

\begin{proof}
	Suppose $w$ is a full run without $e$ in $\ffirst(w)$.
	Since $e$ is always possible until $\be$ is executed, we must have $\be$ on
	$w$.
	This must be preceded by $b$.
	So the first part, call it $u$, of $w$ consists of synchronizations of clients with $S_r$,
	without $S_l$ moving until $S_r$ completes its run, that is until it does $b$.
	From the form of $S_r$ it follows that the first part of $u$ is a sequence
	$u_1\dots u_n$  where each $u_i$ is either $\th_i$ or $\bth_i$.
	This defines a valuation $\val$.
	At this stage, the clients corresponding to literals true in $\val$ are in
	their bottom states where the actions on these literals are possible.
	The other clients are in their top states where the action on the
	corresponding literals are impossible.
	So under our assumption that $S_l$ does not move: $S_r$ can get to action $b$
	iff $\val$ is a satisfying valuation.
	As $\f$ is not SAT this is impossible.
	Hence, such $w$ cannot exist.
\end{proof}

\begin{figure}
	\centering
	\input{figures/fig8.tex}
	\caption{A sound a complete transition system for $\Prog_\f$ when $\f$ is not SAT}
	\label{fig:not-sat-ts}
\end{figure}
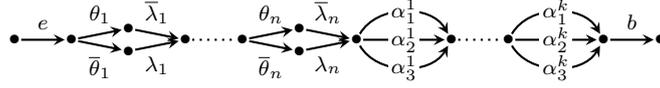

\begin{lemma}\label{lem:P-not-sat}
	If $\f$ is not SAT then the transition system from Figure~\ref{fig:not-sat-ts}
	is a "sound" and "complete" transition system for $\Prog_\f$.
\end{lemma}

\begin{proof}[Proof of Lemma~\ref{lem:P-not-sat}]
	First we check that it is "sound", namely that every "full" path of this
	transition system is a run of $\Prog_\f$.
	For this observe that in the state just before $\a$'s all clients are in their
	bottom states, so all the literals are available.
	This implies that all local paths of $S_r$ on $\a$'s are possible.
	Then action $b$ is possible, but $\be$ is not possible since $S_l$ is in its
	bottom state.

	It remains to verify that the transition system is "complete".
	By Lemma~\ref{lem:only-e}, $e$ is a "first" action of all full runs of $\Prog_\f$.
	Take such a run $ew$.
	Looking at $\Prog_\f$ we see that after $e$ there are four possible actions: $\l_1,\bl_1,\theta_1,\bth_1$.
	We will consider only the case when the next action is $\theta_1$ as the argument
	for the other possibilities is analogous.
	So $w=\theta_1 u_1 \bl_1 u'_1$ for some sequences $u_1$ and $u'_1$;
	observe that $\bl_1$ must appear on the run, as after $e\theta_1$ there is no
	action that can disable it.
	Since the only action on which $S_l$ can synchronize is $\bl_1$, all actions
	in $u_1$ are synchronizations with $S_r$.
	Clearly they do not involve $\bC_1$, as there is no way to execute $\bth_1$ at
	this stage.
	Hence, $u_1$ is independent of $\bl_1$ giving us that $w$ is "trace equivalent"
	to $\theta_1\bl_1u_1u'_1$.
	Repeating this reasoning we obtain that $w$ is trace equivalent to a run in
	the transition system from Figure~\ref{fig:not-sat-ts}.
\end{proof}

\begin{lemma}\label{lem:sat-then-big}
	If $\f$ is SAT then there are runs containing $\be$.
	In every sound and complete reduced transition system for $\Prog_\f$ there are at least
	as many states as there are satisfying valuations for $\f$.
\end{lemma}

\begin{proof}
	For a valuation $\val$ we consider a run $w_\val$ taking $\theta_i$ if
	$\val(x_i)=\true$ and taking $\bth_i$ if $\val(x_i)=\false$.
	For example, if  $x_1,x_n$ hold and $x_2$ does not hold in $\val$ then this
	run would look something like:
	\begin{equation*}
		\theta_1\bth_2\dots\theta_n\a^1_{i_1}\dots\a^k_{i_k}b\be\bl_1\l_2\dots\bl_n
	\end{equation*}
	Here $\a^j_{i_j}$ is a literal that holds in the clause $j$.
	Since $\val$ is a satisfying valuation server $S_r$ can get till $b$.

	Observe that there is no concurrency in the run $w_\val$.
	All $\th$ and $\bth$ actions synchronize with server $S_r$.
	Then $b$ and $\be$ are also dependent on each other as they happen on the same
	client.
	Action $\be$ is the first action of $S_l$.
	It is  followed by a sequence of actions of $S_l$.

	Consider two different satisfying valuations $\val_1$ and $\val_2$.
	Let $s_1$ be the state of program $\Prog_\f$ reached after $\be$ on the run
	$w_{\val_1}$.
	Similarly, for $s_2$ ad $w_{\val_2}$.
	In state $s_1$, the clients that are in the
	bottom states are those corresponding to literals that are true in $\val_1$
	and similarly for $\val_2$. Since $v_1$ and $v_2$ are distinct valuations,
	the states $s_1$ and $s_2$ are distinct.
	So any sound and complete transition system for $\Prog_\f$ must have at least as
	many states as there are satisfying valuations of $\f$.
\end{proof}

We write $|\f|$ for the length of $\f$.
Observe that this is an upper bound on the number of variables as well as on
the number of clauses in $\f$, namely, $n,k<|\f|$.

\begin{corollary}\label{cor:reduction}
	If $\f$ is not SAT then $"\minTS"(\Prog_\f)\leq 6|\f|$ states.
	If $\f$ is SAT then $"\minTS"(\Prog_\f)$ is bigger than the number of
	satisfying valuations for $\f$.
\end{corollary}

\paragraph{Proof of Theorem~\ref{thm:por-impossible}}
Suppose to the contrary that $\Alg$ is an "excellent POR algorithm".
We use it to solve SAT in deterministic polynomial time.

Let $q(x)$ and $r(x)$ be the polynomials associated to $\Alg$.
Namely, $\Alg$ working in time $r(|\Prog|+|\minTS(\Prog)|)$ produces a "sound" and
"complete" reduced transition system of size at most $q(|\minTS(\Prog)|)$.

Given a formula $\p$, consider an integer $m$ and a formula
\begin{equation*}
	\f\equiv\p\land(z_{1}\lor z_2)\dots\land 	(z_{2m-1}\lor z_{2m})
\end{equation*}
where $z_1,\dots,z_{2m}$ are new variables.
Clearly $\f$ is satisfiable iff $\p$ is.
If $\p$ is satisfiable then $\f$ has at least $2^m$ satisfying valuations.

Now we construct our program $\Prog_\f$ and run $\Alg$ on it for
$r(12|\f|)$
time.
If $\f$ is not SAT then, by Corollary~\ref{cor:reduction}, the algorithm stops
and produces a sound and complete 	transition system.
If $\f$ is SAT then by Corollary~\ref{cor:reduction} the algorithm cannot stop in this time as the smallest
sound and complete transition system for $\Prog_\f$ has at least $2^m$ states,
and we can choose $m$ big enough so that $2^m>r(6|\f|)$.

%% file: figures/fig7.tex
\begin{tikzpicture}[line width=0.8pt,>=stealth,font=\scriptsize]
    \begin{scope}
        \node (C1) at (0,0.35) {$\mathbf{C}_i$};
        \node[inner sep=0pt] (n0) at (0,0) {$\bullet$};
        \node[inner sep=0pt] (n1) at (0,-0.5) {$\bullet$};
        \draw[->] (n0) edge[bend right] node[left] {$\theta_i$} (n1);
        \draw[->] (n0) edge[bend left] node[right] {$\lambda_i$} (n1);
        \draw[->] (n1) edge[loop below] node[below] {$x_i$} (n1);    
    \end{scope}
    \begin{scope}[xshift=1.5cm]
        \node (C1) at (0,0.35) {$\mathbf{\bC}_i$};
        \node[inner sep=0pt] (n0) at (0,0) {$\bullet$};
        \node[inner sep=0pt] (n1) at (0,-0.5) {$\bullet$};
        \draw[->] (n0) edge[bend right] node[left] {$\overline{\theta}_i$} (n1);
        \draw[->] (n0) edge[bend left] node[right] {$\overline{\lambda}_i$} (n1);
        \draw[->] (n1) edge[loop below] node[below] {$\bx_i$} (n1);    
    \end{scope}
    \begin{scope}[xshift=3cm]
        \node (C*) at (0,0.35) {$\mathbf{C^*}$};
        \node[inner sep=0pt] (n0) at (0,0) {$\bullet$};
        \node[inner sep=0pt] (n1) at (0,-0.5) {$\bullet$};
        \draw[->] (n0) edge node[left] {$e$} (n1);
    \end{scope}
    \begin{scope}[xshift=4cm]
        \node (bC*) at (0,0.35) {$\mathbf{\bC^*}$};
        \node[inner sep=0pt] (n0) at (0,0) {$\bullet$};
        \node[inner sep=0pt] (n1) at (0,-0.5) {$\bullet$};
        \node[inner sep=0pt] (n2) at (0,-1) {$\bullet$};
        \draw[->] (n0) edge node[right] {$b$} (n1);
        \draw[->] (n1) edge node[right] {$\be$} (n2);
    \end{scope}
    \begin{scope}[xshift=6cm]
        \begin{scope}
            \node (Sl) at (-0.4,0) {$\mathbf{S_l}$};
            \node[inner sep=0pt] (n0) at (0,0) {$\bullet$};
            \node[inner sep=0pt] (n1) at (0.75,0) {$\bullet$};
            \node[inner sep=0pt] (n2) at (1.5,0) {$\bullet$};
            \node[inner sep=0pt] (n3) at (2.25,0) {$\bullet$};
            \node[inner sep=0pt] (n4) at (3,0) {$\bullet$};
            \draw[->] (n0) edge[bend right] node[below] {$e$} (n1);
            \draw[->] (n0) edge[bend left] node[above] {$\be$} (n1);
            \draw[->] (n1) edge[bend right] node[below] {$\lambda_1$} (n2);
            \draw[->] (n1) edge[bend left] node[above] {$\overline{\lambda_1}$} (n2);
            \draw[dotted] (n2) -- (n3);
            \draw[->] (n3) edge[bend right] node[below] {$\lambda_n$} (n4);
            \draw[->] (n3) edge[bend left] node[above] {$\overline{\lambda_n}$} (n4);
        \end{scope}
        \begin{scope}[yshift=-1.1cm]
            \node (Sr) at (-0.4,0) {$\mathbf{S_r}$};
            \node[inner sep=0pt] (n0) at (0,0) {$\bullet$};
            \node[inner sep=0pt] (n1) at (0.75,0) {$\bullet$};
            \node[inner sep=0pt] (n2) at (1.5,0) {$\bullet$};
            \node[inner sep=0pt] (n3) at (2.25,0) {$\bullet$};
            \node[inner sep=0pt] (n4) at (3.5,0) {$\bullet$};
            \node[inner sep=0pt] (n5) at (4.25,0) {$\bullet$};
            \node[inner sep=0pt] (n6) at (5.5,0) {$\bullet$};
            \node[inner sep=0pt] (n7) at (6.25,0) {$\bullet$};
            \draw[->] (n0) edge[bend right] node[below] {$\theta_1$} (n1);
            \draw[->] (n0) edge[bend left] node[above] {$\overline{\theta_1}$} (n1);
            \draw[dotted] (n1) -- (n2);
            \draw[->] (n2) edge[bend right] node[below] {$\theta_n$} (n3);
            \draw[->] (n2) edge[bend left] node[above] {$\overline{\theta_n}$} (n3);
            \draw[->] (n3) edge[bend right=60] node[pos=0.5,fill=white,inner sep=1pt] {$\alpha_1^1$} (n4);
            \draw[->] (n3) edge node[pos=0.5,fill=white,inner sep=1pt] {$\alpha_2^1$} (n4);
            \draw[->] (n3) edge[bend left=60] node[pos=0.5,fill=white,inner sep=1pt] {$\alpha_3^1$} (n4);
            \draw[dotted] (n4) -- (n5);
            \draw[->] (n5) edge[bend right=60] node[pos=0.5,fill=white,inner sep=1pt] {$\alpha_1^k$} (n6);
            \draw[->] (n5) edge node[pos=0.5,fill=white,inner sep=1pt] {$\alpha_2^k$} (n6);
            \draw[->] (n5) edge[bend left=60] node[pos=0.5,fill=white,inner sep=1pt] {$\alpha_3^k$} (n6);
            \draw[->] (n6) edge node[above] {$b$} (n7);
        \end{scope}
    \end{scope}
\end{tikzpicture}

%% file: figures/fig8.tex
\begin{tikzpicture}[line width=0.8pt,>=stealth,font=\scriptsize]
    \node[inner sep=0pt] (n0) at (0,0) {$\bullet$};
    \node[inner sep=0pt] (n1) at (0.75,0) {$\bullet$};
    \node[inner sep=0pt] (n2) at (1.5,0.15) {$\bullet$};
    \node[inner sep=0pt] (n3) at (1.5,-0.15) {$\bullet$};
    \node[inner sep=0pt] (n4) at (2.25,0) {$\bullet$};
    \draw[->] (n0) edge node[above]{$e$} (n1);
    \draw[->] (n1) edge node[above]{$\theta_1$} (n2);
    \draw[->] (n1) edge node[below]{$\overline{\theta}_1$} (n3);
    \draw[->] (n2) edge node[above]{$\overline{\lambda}_1$} (n4);
    \draw[->] (n3) edge node[below]{$\lambda_1$} (n4);
    \node[inner sep=0pt] (n5) at (3,0) {$\bullet$};
    \draw[dotted] (n4) -- (n5);
    \node[inner sep=0pt] (n6) at (3.75,0.15) {$\bullet$};
    \node[inner sep=0pt] (n7) at (3.75,-0.15) {$\bullet$};
    \node[inner sep=0pt] (n8) at (4.5,0) {$\bullet$};
    \draw[->] (n5) edge node[above]{$\theta_n$} (n6);
    \draw[->] (n5) edge node[below]{$\overline{\theta}_n$} (n7);
    \draw[->] (n6) edge node[above]{$\overline{\lambda}_n$} (n8);
    \draw[->] (n7) edge node[below]{$\lambda_n$} (n8);
    \node[inner sep=0pt] (n9) at (5.75,0) {$\bullet$};
    \draw[->] (n8) edge[bend left=60] node[pos=.5,fill=white,inner sep=1pt] {$\alpha_1^1$} (n9);
    \draw[->] (n8) edge node[pos=.5,fill=white,inner sep=1pt] {$\alpha_2^1$} (n9);
    \draw[->] (n8) edge[bend right=60] node[pos=.5,fill=white,inner sep=1pt] {$\alpha_3^1$} (n9);
    \node[inner sep=0pt] (n10) at (6.5,0) {$\bullet$};
    \draw[dotted] (n9) -- (n10);
    \node[inner sep=0pt] (n11) at (7.75,0) {$\bullet$};
    \draw[->] (n10) edge[bend left=60] node[pos=.5,fill=white,inner sep=1pt] {$\alpha_1^k$} (n11);
    \draw[->] (n10) edge node[pos=.5,fill=white,inner sep=1pt] {$\alpha_2^k$} (n11);
    \draw[->] (n10) edge[bend right=60] node[pos=.5,fill=white,inner sep=1pt] {$\alpha_3^k$} (n11);
    \node[inner sep=0pt] (n12) at (8.5,0) {$\bullet$};
    \draw[->] (n11) edge node[above] {$b$} (n12);
\end{tikzpicture}

%% file: pifs.tex
\section{Approximating $\IFS$ test}\label{sec:pifs}

In Listing~\ref{lst:lex}, we have seen a simple and relatively promising algorithm for partial-order reduction based on an $"\IFS"$ oracle.
We have also seen that the oracle test is NP-complete (Proposition~\ref{prop:ifs-is-hard}).
This was followed by a much stronger negative result, stating that we cannot hope for a polynomial-time algorithm guaranteeing to produce close to optimal size "reduced transition systems" (Theorem~\ref{thm:por-impossible}).
We interpret the later result as an indication that $\IFS$ based algorithm is worth serious reconsideration, particularly because it uses clear concepts.
The main challenge is how to approximate $\IFS$ test.

Here we develop heuristics for $"\IFS"(s,B)$ test.
For our algorithms to be correct, we need to ensure one-sided error, namely, a heuristic must say 'yes', if $\IFS(s,B)$ holds.
So the heuristic of always answering 'yes' is a possibility, just not a very attractive one.

We start with a characterization of when $\IFS(s,B)$ holds.
This characterization is, unavoidably, computationally difficult.
We get heuristics for $\IFS$ by weakening the conditions in the
characterization.

Let us fix in this section a "client/server system" $"\Prog"$ over a set of
processes $\Proc=\Clients\cup\Servers$.
For every process $p\in\Proc$ we have a "transition system"
$\TS_p=\struct{S_p,\Sigma_p,s_p^0,\set{\act{a}}_{a\in \Sigma_p}}$.
These determine the transition system $"\TS(\Prog)"$ whose states $s$ are tuples
of states $\set{s_p}_{p\in\Proc}$ and actions are the actions of the individual
processes $\Sigma=\bigcup_{p\in\Proc}\Sigma_p$.
Recall that we write $s_p\act{a}_p$ for local transitions in $\TS_p$, for
process $p$.
We write $s\act{a}$ for global transitions in $\TS(\Prog)$.
Moreover, we use $"\enabled"(s)$ for the set of
labels of transitions outgoing from the global state $s$ of $\TS(\Prog)$.
Recall that $"\dom"(b)$ is the set of processes in $\Prog$ using action $b$.
We extend this notation to sequences and sets of actions: $\dom(v)$ and
$\dom(B)$ denote the unions of domains of the involved actions.

\subsection*{A characterization of $\IFS(s, B)$}

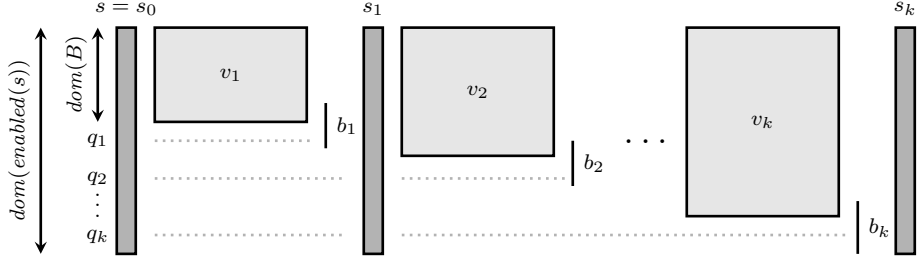
\begin{figure}
  \centering
  \input{figures/fig9.tex}
  \caption{Conditions of Lemma~\ref{lem:ifs}}
  \label{fig:IFS}
\end{figure}

The conditions from the characterization lemma below are schematically
represented in Figure~\ref{fig:IFS}.
We want a pattern ``proving'' that there is a run $u$ from a state $s$ with
$"\first"(u)\incl B$.
Observe that all actions in $\first(u)$ are enabled in $s$. In order to get $\first(u) \incl B$, we are thus looking for a run $u$ where every action $a$ enabled in $s$ which is not in $B$, is preceded by an action $b$ such that $\dom(a) \cap \dom(b) \neq \es$.
As shown in Figure~\ref{fig:IFS}, we are looking for a ``staircase'' of "sticking actions" $b_i$ that block all "enabled" actions in $s$.
Thus the run $u$ should start with a prefix $v_1$ which has its domain included in $B$.
Then $b_1$ is an action which involves a process appearing on $v_1$ but also using a new process $q_1$ not in $\dom(B)$. Observe that thanks to $b_1$, no action involving process $q_1$ is in $\first(u)$.
After $b_1$, $u$ contains a sequence of actions $v_2$ with their domain in $\dom(B) \cup \{q_1\}$, followed by an action $b_2$ that involves a process in $\dom(B) \cup \{q_1\}$ and a new process $q_2$, and so on.
The pattern $v_1b_1v_2b_2\dots v_kb_k$ from Figure~\ref{fig:IFS} and Lemma~\ref{lem:ifs} below may not be by itself a "maximal run", but every maximal run prolonging it will have its $"\ffirst"$-set
included in $B$.
Observe that the conditions guarantee $v_1$ being not empty.

\begin{definition}
  Let $s$ be a global state, and let $R$ be a set of processes.
  We say that an action $c$ ""sticks"" from $R$ in $s$ if $"\dom"(c)=\set{p,q}$ for some
  $p\in R$, $q\not\in R$, and moreover $s_q\act{c}_q$.

  We say that $R$ \emph{wraps $"\enabled"(s)$} if for every $e\in"\enabled"(s)$,
  $"\dom"(e)\cap R\not=\es$.
\end{definition}

\begin{lemma}\label{lem:ifs}
  There is a "maximal" "run" $u$ from $s$ with $\ffirst(u)\incl B$, if and only if,
  there is a "run" from $s$ of the form $v_1b_1v_2b_2\dots v_kb_k$, such that for
  $i=1,\dots,k$ we have:
  \begin{itemize}
    \item $\dom(v_1)\incl \dom(B)$;
    \item $b_i$ sticks from $\dom(v_1b_1\dots v_{i-1}b_{i-1}v_{i})$;
    \item $\dom(v_1b_1\dots v_ib_iv_{i+1})=\dom(v_1b_1\dots
            v_{i}b_{i})$;
    \item $\dom(v_1b_1\dots v_{k}b_k)$ "wraps" $"\enabled"(s)$.
  \end{itemize}
\end{lemma}

\begin{proof}
  For the right-to-left implication, consider a run $u=v_1b_1\dots v_kb_k$ from
  $s$ as in the statement of the lemma.
  Observe that $\ffirst(u)=\ffirst(v_1)\incl B$ and for every $e\in\enabled(s)$,
  $\dom(e)\cap\dom(u)\not=\es$.
  This is because $\dom(b_i)\cap\dom(v_1b_1\dots b_{i-1}v_i) \neq \es$.
  Let $w$ be a complete run prolonging $u$.
  As necessarily $\ffirst(w)\incl\enabled(s)$, we have $\ffirst(w)=\ffirst(u)$.

  For the left-to-right direction take a "maximal run" $w$ with
  $"\ffirst"(w)\incl B$.
  Consider a "trace equivalent" "run" $v_1b_1w_2"\sim" w$ where $v_1$ is the longest
  trace prefix of $w$ such that $"\dom"(v_1)\incl "\dom"(B)$.
  Considering $b_1$, we must have $b_1D v_1$, so there is $p_1\in\dom(b_1)\cap
    \dom(v_1)$.
  On the other hand $\dom(b_1)\not\incl B$ so there is $q_1\in\dom(b_1)-\dom(v_1)$.
  This means that $s_{q_1}\act{b_1}_{q_1}$.
  So $b_1$ is sticking out from $\dom(v_1)$.

  For the remaining part of the "run" $w_2$, we have $"\ffirst"(w_2)\incl\dom(v_1)=\dom(v_1b_1)$.
  We consider a "trace equivalent" "run" $w_2"\sim" v_2b_2w_3$ with $v_2$ the longest
  trace prefix with $"\dom"(v_2)\incl \dom(v_1b_1)$.
  A similar reasoning as for $b_1$, gives us that $b_2$ sticks from $\dom(v_1b_1v_2)$.
  Continuing like this we exhaust all the run $w$.
  If for the final $R=\dom(v_1b_1\dots v_kb_k)$ there would still be $a\in\enabled(s)$ with
  $\dom(a)\cap R=\es$ then $wa$ would also be a run from $s$, that is
  impossible since we assumed $w$ to be maximal.
\end{proof}

\subsection{Approximating $\IFS$ with $\PIFS$}

The condition of Lemma~\ref{lem:ifs} is difficult to check algorithmically because
it involves checking the existence of global runs $v_1,\dots,v_k$.
As a  heuristic we propose a weakening of this condition by approximating the existence of such runs
by local runs, namely runs in transition systems of respective processes.

\begin{definition}\label{def:pifs}\label{def:rpifs}
  A set of actions $B$ ""potentially includes a first set"" in $s$,
  denoted $"\PIFS"(s,B)$, if there is a sequence of sets of actions
  $B=B_0\incl B_1 \incl \dots \incl  B_k$ such that:
  \begin{itemize}
    \item $B_{i+1}=B_i\cup C_i$ , where $C_i$ is the set of actions $c$ for which
          there is a process $p$ and an action $b\in B_i$, with $c$ sticking out from $\set{p}$ in
          $s$, and a local path $s_p\act{b}_p\act{x}_p\act{c}_p$, for some $x$ such
          that $"\dom"(x)\incl "\dom"(B_i)$;
    \item $"\dom"(B_k)$ wraps $"\enabled"(s)$.
  \end{itemize}
  A set of actions $B$ ""remotely potentially includes first"" set in $s$,
  denoted $""\rPIFS""(s,B)$, if there is a sequence $B=B_0\incl B_1 \incl \dots
    \incl  B_k$ satisfying the same two conditions without the clause $"\dom"(x)\incl "\dom"(B_i)$.
\end{definition}

\begin{lemma}\label{lem:pifs}
  For every $s$ and $B$: $"\IFS"(s,B)$ implies $"\PIFS"(s,B)$ that in turn
  implies $"\rPIFS"(s,B)$.
\end{lemma}

\begin{proof}[Proof of Lemma~\ref{lem:pifs}]
  Consider a run $v_1b_1\dots v_kb_k$ from the characterization of $\IFS(s,B)$
  in Lemma~\ref{lem:ifs}.
  We show that every $b_i$ belongs to some $B_j$ from the definition of $\PIFS$.
  Consider $b_1$ that "sticks" from $\dom(v_1)$ in $s$.
  This means $\dom(b_1)=\set{p_1,q_1}$ with $p_1\in\dom(v_1)$,
  $q_1\not\in\dom(v_1)$ and $s_{q_1}\act{b_1}_{q_1}$.
  Consider the first action $a_1$ of $p_1$ on $v_1=v'_1a_1v''_1$.
  We have $s_{p_1}\act{a_1}_{p_1}\act{x}_{p_1}\act{b_1}_{p_1}$ for some $x$ with $\dom(x)\incl
    B_1$.
  If $a_1\in\ffirst(v_1)$ then $b_1\in B_1$.
  Otherwise, $\dom(a_1)=\set{p_1,r_1}$, and we repeat the reasoning for $r_1$
  instead of $p_1$.
  An induction gives us $a_1\in B_l$ for some $l$; and in
  consequence   $b_1\in B_{l+1}$.
  The argument for all other $b_i$'s is similar.

  Now, consider some $e\in"\enabled"(s)$.
  By definition of $\IFS(s,B)$ we have $"\dom"(e)\cap \dom(v_1b_1\dots v_kb_k)\not=\es$.
  But $\dom(v_1b_1\dots v_kb_k)=\dom(v_1)\cup\set{q_1,\dots,q_k}\incl
    \dom(B)\cup\dom(\set{b_1,\dots,b_k})\incl \dom(B_k)$.
  So the condition of $\dom(B_k)$ wrapping $\enabled(s)$ is satisfied.
\end{proof}

The algorithm for $\PIFS(s,B)$ is presented in Listing~\ref{lst:pifs}.
It follows rather directly the formula from Definition~\ref{def:pifs}.
In each iteration of while-loop, $B$ is extended with $C_i$ computed by the two
for-loops.

\begin{lstlisting}[caption={Computing $\PIFS$},label=lst:pifs,frame=lines,columns=fullflexible]
procedure $\PIFS(s,B)$:
  $D:=\es$
  while $B-D\not=\es$ and not($\dom(B)$ wraps $\enabled(s)$)
    $D:=B$
    forall $q$, $b$ such that $s_q\act{b}$
      let $p\in\dom(b)$, $p\not=q$
        if $\exists d\in D.\ s_p\act{d}_p\act{x}_p\act{b}_p$ with $x \in \dom(B)$£\label{ln:test}£
          add $b$ to $B$
  return($\dom(B)$ wraps $\enabled(s)$)
\end{lstlisting}

The test in line~\ref{ln:test} asks if there is a local "run" from $s_p$ in the
transition system $"\TS"_p$ of process $p$ of the form $dxb$ where $x$ is a sequence of
actions such that $"\dom"(x)\incl "\dom"(B)$.
In order to compute $\rPIFS$ instead of $\PIFS$ it is enough to
remove the requirement $x \in \dom(B)$, that is the requirement on  actions over which $b$ is reachable.

The reason to introduce a less precise $\rPIFS$ is that the test of
``$s_p\act{b}_p\act{x}_p\act{c}_p$, for some $x$ such
that $"\dom"(x)\incl "\dom"(B_i)$'' in the definition of $\PIFS$ may be expensive in some situations.
If this test were implemented by a procedure computing
reachability each time, then indeed the resulting exploration algorithm would
spend most of its running time doing this test.
Our preferred solution is to precompute, for every process $p$, all triples
$(s_p,R,b)$ such that $s_p\act{x}_p\act{b}_p$ for some $x$ with $\dom(x)\incl R$.
With this set of triples the test becomes very fast.
It is of course not difficult to come up with a transition system for which the set
of such triples is exponential in the size of the transition system.
Yet, if the transition system of a process is a tree then the number of these
triples is linear in the size of the transition system.
In our examples, even if processes are relatively big DAGs, the number of triples
is never a problem.
Similarly, for servers that are not even DAGs.
For servers implementing variables, locks, or communication channels, the number
of these triples is linear in the size of the server.

\begin{remark}\label{rem:reversals}
Of course, the above approximation of $"\IFS"$ is not the only one possible.
A possible idea is to approximate the existence of a global run by runs of
$k$-processes instead of a single process, similarly to $k$-Cartesian abstraction~\cite{Hoe.Maj.Pod.Thread2017}.
Another option is the race reversal technique introduced in~\cite{Abd.Aro.Jon.Sag.Source2017}.
Let us briefly sketch it below.

Suppose we do not look for the existence of the whole pattern from Figure~\ref{fig:IFS} but only for the first part $v_1b_1$. If we cannot find $v_1b_1$ part, then clearly we cannot find the whole pattern either.
Assume that we use an exploration algorithm like the one from
Listing~\ref{lst:lex}.
Consider an execution that comes back to node $n$, after finishing exploring $n_a$ with $n\act{a}n_a$.
At this point, we may ask if $a$ is a first action of every run from the state $s(n)$; if so then we can stop exploring from $n$.
In other words, we would like to know whether $\IFS(s,\enabled(n)-\set{a})$ holds.
To approximate this, we may ask if there is a path $u_1b_1$ from $n_a$ such that $u_1\sim v_1b_1w_1$ with $\dom(v_1)\cap\dom(a)=\es$ and $b_1 D a$.
This is exactly to say that there is a race between $a$ and $b_1$.
But this is also exactly the criterion for the first part $v_1b_1$ in the $\IFS$
pattern from the characterization in Lemma~\ref{lem:ifs}.
Thus, the race reversal technique can be seen as another way of approximating $\IFS$ test.

There are two complications with implementing this technique in our setting.
First, because we allow blocking, the "sleep-invariant" is not strong enough.
An exploration algorithm must guarantee that every ``sleep-maximal'' run is
explored:
\begin{quote}
after exploration of a node $n$: for every run $u$ from $s(n)$ such that there is
no longer run $ub$ with $\ffirst(ub)\cap\Sleep(n)=\es$ we have a path $v$ from
$n$ with $u\sim v$.
\end{quote}
This implies that the algorithm may potentially explore more.
The other, more serious, problem is how to find a race pattern.
In stateless exploration, it is simple, as it is enough to go up the exploration path.
In our case, this requires going up the exploration DAG.
But the number of paths may be exponential in the size of a DAG.
Theoretically, this race test becomes NP-complete, but we may still hope to have
efficient heuristics.

\end{remark}

%% file: figures/fig9.tex
\begin{tikzpicture}[line width=1pt,>=stealth,font=\scriptsize]
    \draw[fill=black!30] (0,0) rectangle (0.25,-3);
    \draw (0.125,0.25) node {$s=s_0$};
    \draw[<->] (-0.25,-1.25) -- node[above,sloped] {$\dom(B)$} (-0.25,0);
    \draw (-0.25, -1.5) node {$q_1$};
    \draw (-0.25, -2) node {$q_2$};
    \draw (-0.25, -2.25) node {$\vdots$};
    \draw (-0.25, -3) node[above] {$q_k$};
    \draw[<->] (-1,-3) -- node[above,sloped] {$\dom(\enabled(s))$} (-1,0);
    \draw[fill=black!10] (0.5,0) rectangle node {$v_1$} (2.5,-1.25);
    \draw (2.75,-1) edge node[right] {$b_1$} (2.75,-1.6);
    \draw[black!30,dotted] (0.5,-1.5) -- (2.6,-1.5);
    \draw[fill=black!30] (3.25,0) rectangle (3.5,-3);
    \draw (3.375,0.25) node {$s_1$};
    \draw[fill=black!10] (3.75,0) rectangle node {$v_2$} (5.75,-1.7);
    \draw (6,-1.5) edge node[right] {$b_2$} (6,-2.1);
    \draw[black!30,dotted] (0.5,-2) -- (3,-2);
    \draw[black!30,dotted] (3.75,-2) -- (5.9,-2);
    \draw (7,-1.5) node {{\Large $\dots$}};
    \draw[fill=black!10] (7.5,0) rectangle node {$v_k$} (9.5,-2.5);
    \draw (9.75,-2.3) edge node[right] {$b_k$} (9.75,-3);
    \draw[black!30,dotted] (0.5,-2.75) -- (3,-2.75);
    \draw[black!30,dotted] (3.75,-2.75) -- (9.65,-2.75);
    \draw[fill=black!30] (10.25,0) rectangle (10.5,-3);
    \draw (10.375,0.25) node {$s_k$};
\end{tikzpicture}

%% file: explore.tex
\section{A new partial-order reduction algorithm}\label{sec:algo}

Our starting point is the algorithm from Listing~\ref{lst:lex} employing $\IFS$
oracle to avoid exploration of branches leading to "sleep-blocked" states.
We intend to use $\PIFS$ approximation instead.
Since this is only an approximation, we will add two ingredients to improve the performance.
The first is a heuristic to compute some "covering source set".
Interestingly, we derive it as some kind of dual to $\rPIFS$ test.
The second ingredient is a function for choosing the most promising action to
explore among those in the computed source set.
The idea is to choose actions in such a way that the $\PIFS$ test becomes false as soon as possible.

\subsection{A $\rPIFS$ based $\closure$ operation}\label{sec:closure}
We come back to the notion of "covering source set" from Section~\ref{sec:por}.
Until now, we have not given any algorithm to compute such sets.
We will derive one now from $"\rPIFS"$ test (Definition~\ref{def:rpifs}).
We will then compare it to a method of computing "persistent" sets, and show
that $\rPIFS$-based source sets can be smaller than persistent sets.

\begin{definition}
  For a state $s$ and an action $b$, $""\closure""(s,b)$ is the smallest set of
  actions $C$ such that
  \begin{itemize}
    \item $C$ contains $\set{a : \exists p\in\dom(b).\ s_p\act{a}_p}$;
    \item for every $b\in C$, if
          $"\dom"(b)=\set{p,q}$, $s_p\act{b}_p$, and $s_q\act{c}_q\act{}^*_q\act{b}_q$
          then $c\in C$.
  \end{itemize}
\end{definition}

Observe that the second condition in this definition is exactly the same as in $"\rPIFS"$
test, but it is used ``in the other direction''.
(The condition for $\rPIFS$ requires that if $c$ is in the set then $b$ should
be in the set.)
While there is a direct proof of the following lemma,
we prefer to
give here an argument relying on this connection.

\begin{lemma}\label{lem:closure-is-covering}
  If $b\in\enabled(s)$ then $\closure(s,b)\cap\enabled(s)$ is a "covering source set".
\end{lemma}

\begin{proof}
  Consider a "maximal" run $s\act{u}$ from $s$.
  Suppose for the contradiction that $"\first"(u)\cap"\closure"(s, b)=\es$.
  By Lemmas~\ref{lem:ifs} and \ref{lem:pifs}, this means $"\rPIFS"(s,\first(u))$ holds.
  Recalling its definition, this means that there is a sequence of sets
  of actions $B_0,B_1,\dots,B_k$ such that $B_0=\first(u)$ and $B_k$ "wraps" $\enabled(s)$.
  In particular, $B_k\cap\dom(b)\not=\es$.
  By the first condition on $"\closure"$, $B_k\cap \closure(s,b)\not=\es$.
  Say $d_k$ is in this intersection.
  But then by the second condition on $\closure$ and the definition of $\rPIFS$, there must
  be some $d_{k-1}$ in $B_{k-1}\cap \closure(s,b)$.
  Repeating this argument, we get some $d_0\in B_0\cap\closure(s,b)$.
  But $B_0=\first(u)$, a contradiction.
\end{proof}

Let us compare $"\closure"$ to some standard method for
computing a persistent set.
We take the one from~\cite{Far.Klu.Pod.Sound2022} that is the best we could find in the
literature, at least for our model of distributed systems.

\begin{definition}\label{def:pset}
  For an action $b$ and state $s$, a $""\pclosure""(s,b)$ is the smallest set of
  processes $R$ such that
  \begin{itemize}
    \item $\dom(b)\incl R$;
    \item if $p\in R$ and $s_p\act{\ }^*_p\act{c}_p$ then $\dom(c)\incl R$.
  \end{itemize}
  Now, we define $""\pset""(s,b)=\set{a\in \enabled(s) : \dom(a)\incl \pclosure(s,b)}$.
\end{definition}

\begin{remark}
  Comparing the two definitions we get $\closure(s,b)\incl \pset(s,b)$.
  This is because $"\pclosure"(s,b)$ is at least as big as
  $\dom(\closure(s,b))$, and $\pset(s,b)$ contains all the actions whose domain is
  included in $\pclosure(s,b)$.
\end{remark}

\begin{lemma}
  If $b\in\enabled(s)$ then $"\pset"(s,b)$ is a persistent set.
\end{lemma}
\begin{proof}
  For a contradiction, suppose there is a run $s\xrightarrow{a_1\dots a_kc}$ such that
  for all $i=1,\dots,k$ $a_i I \pset(s,b)$, and $c\not\in\pset(s,b)$, but $c
    D\pset(s,b)$.
  So there is $d\in\pset(s,b)$ with $d D c$.
  Let us take $p\in\dom(d)\cap\dom(c)$.
  Since $d\in\pset(s,b)$, we have  $p\in"\pclosure"(s,b)$.
  Observe that  $s_p\act{x}_p\act{c}_p$ for $x$ some subsequence of $a_1\dots a_k$.
  Hence, $\dom(c)\incl \pclosure(s,b)$, and in consequence $c\in\pset(s,b)$.
  A contradiction.
\end{proof}

\begin{remark}
  It is not the case that $\closure(s,b)\cap\enabled(s)$ is always a "persistent" set in $s$.
  Consider the example in Figure~\ref{fig:cs-example} again.
  Action $b$ is enabled in the initial state $s_0$.
  We have $\closure(s_0,b)=\set{b,a,e}$.
  But because of $s_0\act{c}$, the set $\closure(s_0,b)\cap\enabled(s_0)$  set is not "persistent" since $c
    D\closure(s_0,b)$ because of $e$, but $c\not\in\closure(s_0,b)$.
\end{remark}

From this discussion it follows that $"\closure"(s,b)$ is preferable to
$"\pset"(s,b)$, because it is always a covering source set, and smaller than $\pset(s,b)$.
The remaining question is how to choose $b$.
We see two interesting options: (i) ""lex-closure"" is to take the smallest $b$, in our
fixed ordering on actions, among those enabled  in $s$; or (ii) ""min-closure""
is to take $b$ giving the
smallest covering source set in $s$.
The first option may be interesting when not using sleep sets, the second seems
preferable overall.
Simple algorithms for the two options are presented in Listing~\ref{lst:closure}.

\begin{lstlisting}[caption={Two possibilities for computing a closure},label=lst:closure,frame=lines,columns=fullflexible]
  function $\minclosure(n)$ // returns a covering source set
      $C$ := $\enabled(n)-\Sleep(n)$
      forall $b\in \enabled(n)-\Sleep(n)$
        $B$ := $\closure(s(n),b)\cap(\enabled(n)-Sleep(n))$
        if $|B| < |C|$
          $C$ := $B$
      return $C$
    
  function $\lexclosure(n)$ // returns a covering source set
    if $\enabled(n)=\es$ 
      return $\es$
    else
      choose minimal $b\in \enabled(n)$ w.r.t. linear ordering on actions
      return $\closure(s(n),b)\cap(\enabled(n)-\Sleep(n))$
\end{lstlisting}

\subsection{Choose an action from a closure}\label{sec:chooseaction}

The goal of $\chooseaction(s,A)$ is to pick the most promising action from a set
$A$.
One simple implementation is to choose the smallest action in $A$ according to our fixed ordering.
Another option is to use information given by $\PIFS$; we describe it now. 

As we have seen, $\closure$ corresponds to $\rPIFS$.
Since $\PIFS$ is stronger than $\rPIFS$, the question arises if we can compute
some variant of a closure operation corresponding to $\PIFS$.
The challenge here is that while $\closure$ and $\rPIFS$ refer to simple
reachability, $\PIFS$ refers to reachability over a particular set of
actions, and this set of actions is calculated dynamically in $\PIFS$.
We will use this calculated set to decide which action to explore next.
So instead of aiming at a closure corresponding to $\PIFS$, we aim at choosing an
action to explore that is most promising to make the $\PIFS$ test false and stop
exploration.

The idea of $\chooseaction(s,A)$ is to compute $\PIFS(s,\set{b})$ for each
action $b\in A$, and look at the obtained set $B$.
If $B$ "wraps" all actions enabled in $s$ then $\PIFS$ indicates that there may
be a "run" from $s$ with $\set{b}$ as the only first action.
In this case it is certain that we should explore $b$ from $s$.
If $B$ does not wrap all enabled actions then we take $b$ giving the biggest $B$.
The intuition is that this $b$ will be present in $"\ffirst"$ sets of a big portion
of "runs" from $s$.
To implement this idea we slightly modify $\PIFS$ function, so that it returns
the computed set $B$.
The algorithms for the two options are presented in Listing~\ref{lst:choose-action}.

\begin{lstlisting}[caption={Choose action},label=lst:choose-action,frame=lines,columns=fullflexible]
  function $\chooseaction(s,A)$
    $max:=0$
    forall $b\in A$
      $(res,B):=a\PIFS(s,\set{b})$
      if $res$ then return $b$
      if $|B|>max$
        $max$ :=$ |B|$
        $bmax$ := $b$
    return $bmax$
  
  function $a\PIFS(s,B)$:
    $D:=\es$
    while $B-D\not=\es$ and not($\dom(B)$ wraps $\enabled(s)$)
      $D:=B$
      forall $q$, $b$ such that $s_q\act{b}$
        let $p\in\dom(b)$, $p\not=q$
          if $\exists d\in D.\ s_p\act{d}_p\act{x}_p\act{b}_p$ with $dom(x) \subseteq \dom(B)$
            add $b$ to $B$
      if $\dom(B)$ wraps $\enabled(s)$
        return ($\true$,$B$)
      else
        return ($\false,B$)
\end{lstlisting}

\subsection{A $\PIFS$ based algorithm}

As we have seen, $"\closure"$ gives a covering source set, so it can be used to compute what to explore, while
$"\IFS"$ can be used to test when to stop exploring.
We have also seen that if we had a perfect $"\IFS"$ oracle then we would not need
$\closure$.
Of course, dually, if we had a perfect variant of closure operator giving us a minimal
covering "source set" then we would not need $\IFS$.
We have seen in Section~\ref{sec:ifs-hard} that perfect $\IFS$ test is \NP-hard.
For the same reason a perfect closure operator is also \NP-hard.

The idea of the algorithm presented in Listing~\ref{lst:full} is to use the best we have.
For deciding when to stop we use $"\PIFS"$ that is the best approximation of
$\IFS$ we have.
For limiting the number of actions to consider, we use $\chooseclosure$ as
described in Section~\ref{sec:closure}.
Finally, we use $\chooseaction$ to take the most promising action to explore as
discussed in Section~\ref{sec:chooseaction}.

The algorithm from Listing~\ref{lst:full} builds a "reduced transition system" where each node $n$ is associated a state $s(n)$ of $\TS(\Prog)$ and a "sleep set" $\sleep(n)$.
\begin{lstlisting}[caption={Explore using $\PIFS$, $\chooseclosure$ and $\chooseaction$},label=lst:full,frame=lines,columns=fullflexible,float=ht]
  main($"\Prog"$):
  create node $n^0$ with $s(n^0)=s^0$ and $\Sleep(n^0)=\es$
  $\explored:=\es$ 
  $\Explore(n^0)$

  procedure $\Explore(n)$:
    $Sl:=\Sleep(n)$
    $C:=\chooseclosure(n)$ // $C$ is a "covering source set", $C\incl\enabled(n)$
    while $C-Sl\not=\es$
        $a:=\chooseaction(s(n),C-Sl)$
        let $s'$ such that $s(n)\act{a}s'$
        if $\exists n''\in\explored$ s.t $n'' "\subsum" (s',Sl-Da)$
          add edge $n\act{a}n''$ to the exploration graph
        else if $\PIFS(s',"\enabled"(s')-(Sl-Da))$
          create node $n'=(s',Sl-Da)$
          add edge $n\act{a} n'$
          $\Explore(n')$
        $Sl:=Sl\cup\set{a}$
    add $n$ to $\explored$
\end{lstlisting}
At this point we have all the tools to prove its correctness.
The algorithm is a modification of the one from Listing~\ref{lst:lex} that is
correct by Lemma~\ref{lem:lex-optimal}.
Changing $"\IFS"$ to $"\PIFS"$ preserves correctness by Lemma~\ref{lem:pifs}.
Function $\chooseaction$ does not influence correctness.
Finally, the restriction to $C$ in the condition of the while-loop preserves
correctness thanks to Remark~\ref{rem:use-of-closure} and
Lemma~\ref{lem:closure-is-covering} stating that "closure" gives a "covering
source set".

\begin{theorem}
  Given a "client/server system" $\Prog$, the algorithm from Listing~\ref{lst:full} constructs a "sound" and "complete"
  transition system for $"\TS(\Prog)"$.
\end{theorem}

%% file: experiments.tex
\section{Implementation and experimental results}\label{sec:experiments}

We report on the implementation of the algorithm from Listing~\ref{lst:full}.
We have compared it with what we consider the state-of-the art persistent set method, namely $"\pset"$ from Definition~\ref{def:pset}.

The correctness of an implementation as well as of the reported results is probably
the first question that comes to mind.
Our implementation is in C++. It is based on a system developed for different
purposes adapted to the exploration of ``simple transition systems'' we consider in this work. This tool includes several features, such as custom memory management and optimized "subsumption" checking.
Even if we verified the abstract algorithm from Listing~\ref{lst:full} in Coq,
that we are not capable of, this would still leave a large margin for coding errors.
Instead, we use Proposition~\ref{prop:source-covering} to check that the computed "reduced transition system" is "complete".
To that purpose, we compute the transition system $"\TS(\Prog)"$ and for every state $s$ of it, we compute the $"\first"$ sets of all "maximal" runs from $s$.
Since our systems are acyclic, this can be done bottom-up relatively
efficiently.
Then, we take a reduced transition system and check for all its states that the set of outgoing transitions is a "covering source set" as in Definition~\ref{def:covering-source-set}.
According to Proposition~\ref{prop:source-covering}, this test guarantees that the reduced transition system is "complete".
We have run this test on a large number of randomly generated systems as well as on all the reported examples, up to a size when generating the full transition system $\TS(\Prog)$ is still feasible.
This gives us some confidence in the reported results.

We have compared 6 algorithms, all of them using depth-first exploration.
The first one is the standard reachability algorithm without
partial-order reduction (\texttt{reach}).
The others differ in the use of "sleep sets", $"\PIFS"$, and a choice between
"min-closure" or simple "lex-closure".
We consider one algorithm without sleep sets, but with all other improvements,
namely with $"\PIFS"$ and "min-closure" (\texttt{full-sleep}).
Then we consider four algorithms using sleep sets, together with:
persistent sets (\texttt{pset+sleep}), "min-closure" without $"\PIFS"$
(\texttt{minclosure+sleep}), $"\PIFS"$ without "min-closure"
(\texttt{apifs+sleep}) and, finally, with both $"\PIFS"$ and "min-closure"
(\texttt{full+sleep}).
The last option is precisely the algorithm from Listing~\ref{lst:full} implemented using a stack instead of recursion.
We have experimented with breadth-first and some priority queue exploration ordering, but the results were roughly the same or even worse in the case of
BFS.

We report experimental results on 8 models from the literature and 2 new
benchmarks, one of which are randomly generated models with locks.
The models from the literature are:
``Dinning philosophers'' (\texttt{dp}),  ``Dispatcher''
(\texttt{disp})~\cite{Cot.Pet.Rod.Sou.Quasioptimal2021}, ``MPat'' (\texttt{mpat})~\cite{Cot.Pet.Rod.Sou.Quasioptimal2021} and the program in Figure~1
(\texttt{qo}) from~\cite{Cot.Pet.Rod.Sou.Quasioptimal2021}, then ``Token ring''
(\texttt{tk})~\cite{Val.State}, next a model of Peterson's mutual exclusion
algorithm (\texttt{pet})~\cite{Pet.Myths1981,Val.Stop2017}, and finally ``filesystem'' (\texttt{fs})
and ``lastzero'' (\texttt{lz}) from~\cite{Fla.God.Dynamic2005}. The new
benchmarks are described below.
Model (\texttt{qo}), is an example where race reversal
from~\cite{Abd.Aro.Jon.Sag.Source2017} is exponential~\cite{Cot.Pet.Rod.Sou.Quasioptimal2021}.
Models ``filesystem'' and ``lastzero'' are the classical examples for dynamic
partial-order reduction.

\begin{figure}[t]
      \begin{minipage}{.49\textwidth}
            \centering
            \includegraphics[width=\textwidth]{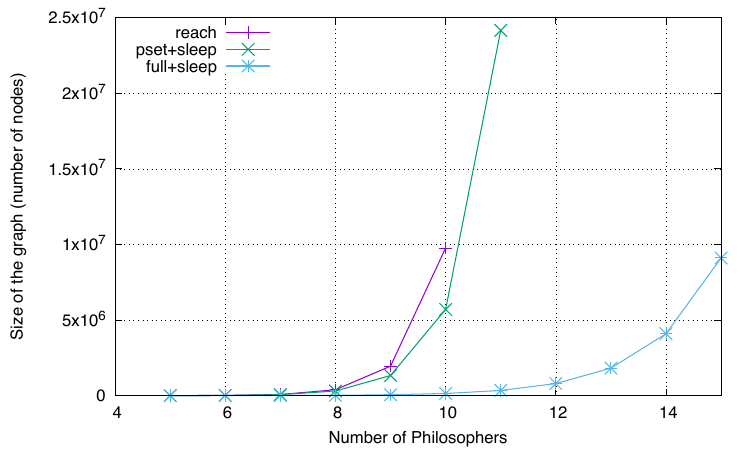}
      \end{minipage}
      \hfill
      \begin{minipage}{.49\textwidth}
            \centering
            \includegraphics[width=\textwidth]{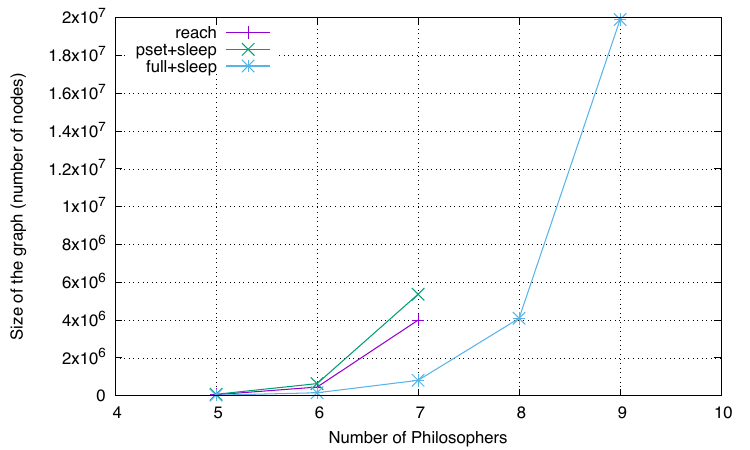}
      \end{minipage}
      \caption{Comparison of algorithms \texttt{reach},
            \texttt{pset+sleep} and
            \texttt{full+sleep} on the ``Dining Philosophers'' model, with the main
            loop unfolded once (left), and twice (right). Experiments have been ran on
            a MacBook Air M2 with 16Gb of memory, and with a timeout of 15min.}\label{fig:graph_dp}
\end{figure}

Figure~\ref{fig:graph_dp} compares the persistent
set algorithm~\texttt{pset+sleep} and our algorithm~\texttt{full+sleep} on the classical
``Dining Philosophers'' example. Each Philosopher is a client process that takes
both forks, eats, and then releases both forks, whereas forks are server
processes that behave as locks.
The picture on the left shows that on this
model, the state-of-the-art persistent set approach (i.e.
\texttt{pset+sleep}) is slightly better than reachability.
Our algorithm \texttt{full+sleep} produces reduced graphs
which are several orders of magnitude smaller than the persistent sets approach,
and can explore  much larger models within a span of
15 minutes (up to $15$
Philosophers instead of only $10$).
The picture on the right of Figure~\ref{fig:graph_dp} compares the same
algorithms on a similar model, but where the Philosophers eat
twice.
This induces much more races in the model, making it harder to explore.
Our algorithm
\texttt{full+sleep} still significantly outperforms the persistent set algorithm.

\begin{figure}[t]
      \begin{minipage}{.49\textwidth}
            \centering
            \includegraphics[width=\textwidth]{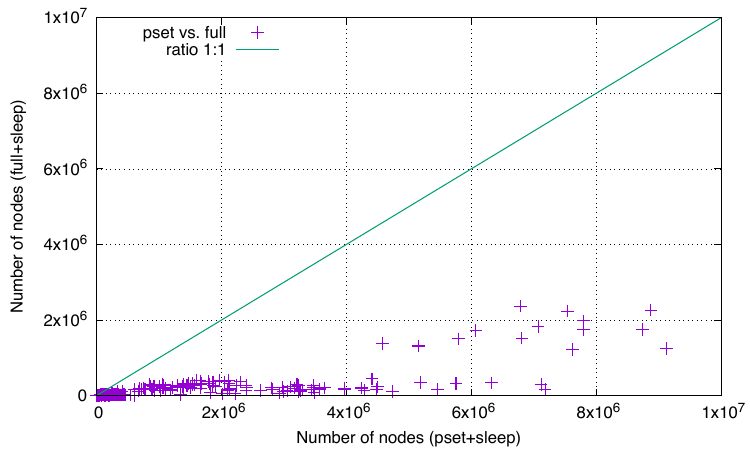}
      \end{minipage}
      \hfill
      \begin{minipage}{.49\textwidth}
            \centering
            \includegraphics[width=\textwidth]{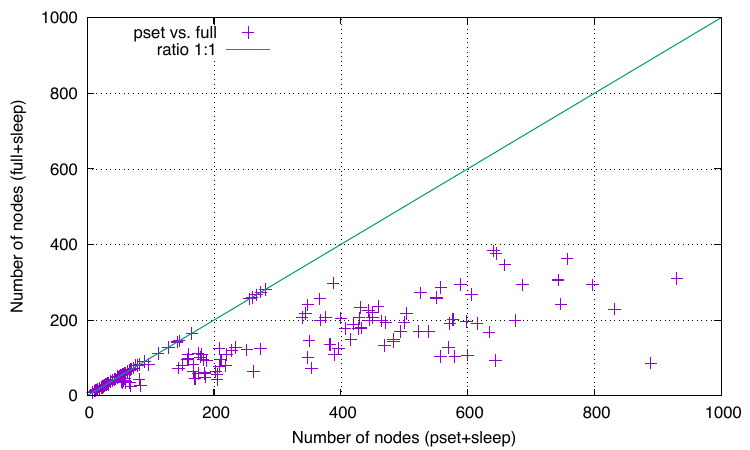}
      \end{minipage}
      \caption{Comparison of algorithms
            \texttt{pset+sleep} and
            \texttt{full+sleep} on the ``Multi locks'' model (full results on the
            left,
            focus on the smaller models on the right). Experiments have been ran on
            a MacBook Air M2 with 16Gb of memory, and with a timeout of
            2min.}\label{fig:graph_ml}
\end{figure}

Figure~\ref{fig:graph_ml} reports a comparison between the persistent sets
approach (\texttt{pset+sleep}) and our approach (\texttt{full+sleep}) on
randomly generated ``Multi locks'' models.
They consist of a set of processes (clients) and a set of locks (servers).
Processes are randomly generated. Each process first acquires a fixed number of
randomly chosen locks, and then releases the locks. We have generated 765
models using 51 random seeds, with either 4, 6, 8, 10 or 12 clients, 10 locks,
and
where each client acquires the same number of locks, which is set to either 1, 2
or 3.
Algorithm \texttt{pset+sleep} could not terminate within 2 minutes on 37
models. The picture on the left of Figure~\ref{fig:graph_ml} compares the size
of the reduced graphs for the remaining 728 models. The picture on the right
focuses on the smaller models with graph of at most 1000 nodes. Both pictures
clearly show a superior performance of \texttt{full+sleep} over
\texttt{pset+sleep}, with gains around an order of magnitude on a
significant share of the models.

\begin{table}[t]
      \centering\tiny
      \input{table-results.tex}
      \caption{Comparison of algorithms \texttt{reach}, \texttt{pset+sleep},
            \texttt{full+sleep}, \texttt{full-sleep}, \texttt{minclosure+sleep} and
            \texttt{apifs+sleep} on various instances of selected models. For each
            instance we indicate the number of clients (\texttt{\#C}) and the number
            of servers (\texttt{\#S}). For each algorithm, we report the number of
            nodes in the computed graphs (\texttt{N}).
            For algorithm \texttt{full+sleep} we also give
            the number of paths (\texttt{{Paths}}) in the reduced model. Experiments
            were performed on MacBook Air M2 with 16Gb of memory, and a timeout of 5min.}
      \label{table:por-reduction}
\end{table}

In Table~\ref{table:por-reduction}, we compare the 6 algorithms on 9
models. We have considered several instances of each model with increasing
number of
processes as indicated along with the model name. The loops contained in the
models have been unfolded once.
The first model in the table, ``boolean gates'' (\texttt{bg}) is a new
model representing an idealized tree architecture.
It consists of a set of gates (clients) that are connected by wires (servers).
The gates are set up in a complete binary tree. The size of the model
corresponds to the height of the tree. Hence, the number of processes increases
exponentially with the height. Each gate reads its two input wires,
then writes its output wire. There are races between connected gates, while
gates on distinct branches are fully concurrent.

For each algorithm and each model, we report the number $N$ of nodes in the
(reduced) graphs.
We first observe that our algorithm \texttt{full+sleep} which includes all the
features presented in this paper clearly outperforms the standard reachability
algorithm (\texttt{reach}) and the persistent sets based partial-order reduction
(\texttt{pset+sleep}), on all the examples.
The other columns allow us to compare
the impact of each technique used in \texttt{full+sleep}.

Comparing \texttt{full+sleep} and \texttt{full-sleep}, we first observe that using
"sleep sets" is beneficial for all models, except \texttt{fs} and \texttt{pet}.
Indeed, in these models, it appears that sleep sets lead to a huge number of state
duplications (nodes with the same program state but different sleep sets).

From columns \texttt{full+sleep}, \texttt{minclosure+sleep} and
\texttt{apifs+sleep}, we see that $"\PIFS"$ does not provide additional benefit over
"min-closure" on the first five models.
Conversely, "min-closure" alone does not improve over $"\PIFS"$ on models
\texttt{lz}, \texttt{tk}, \texttt{pet} and \texttt{fs}. This last model is
particular since most races appear from size $14$. Then \texttt{apifs+sleep} performs significantly better than
\texttt{minclosure+sleep} and even \texttt{full+sleep}.
Finally, the advantages of combining our techniques are clearly visible on
the \texttt{dp} model.

While our techniques scale up significantly better than standard
persistent set based reduction, they are unable to explore bigger instances
of models \texttt{pet} and \texttt{lz}. These models
are indeed out of reach for static partial-order approaches. A combination of
dynamic race detection (which can be seen as one possible heuristic for
$"\IFS"$), as well as coarser "por-equivalences", in place of Mazurkiewicz
"trace equivalence" used here, will permit handling larger instances of
these models.

Finally, the column $Paths$ in Table~\ref{table:por-reduction} shows
the number of paths in the reduced graphs
produced by our best algorithm \texttt{full+sleep}. We can observe that it grows sharply on many examples: \texttt{bg}, \texttt{mpat},
\texttt{tk}, \texttt{pet} and \texttt{dp}.
While our models are probably not optimal in the number of paths these numbers
indicate that a stateless approach might be impractical on these examples.

%% file: table-results.tex
\begin{tabular}{|l r r||r||r||r|r||r||r||r|}
\hline
\multicolumn{3}{|c||}{Models} & \multicolumn{1}{c||}{reach} & \multicolumn{1}{c||}{pset+sleep} & \multicolumn{2}{c||}{full+sleep} & \multicolumn{1}{c||}{full-sleep} & \multicolumn{1}{c||}{minclosure+sleep} & \multicolumn{1}{c|}{apifs+sleep} \\
& \#C & \#S & N  & N  & N & Paths & N & N & N\\
    \hline
bg 3 & 15 & 14 
  & 163217 
  & 95991 
  & \textbf{131} & 256 
  & \textbf{131} 
  & \textbf{131}
  & 329\\
bg 4 & 31 & 30
  & \multicolumn{1}{c||}{timeout}
  & \multicolumn{1}{c||}{timeout}
  & \textbf{401} & 65536
  & \textbf{401}
  & \textbf{401}
  & 5945\\
bg 5 & 63 & 62
  & \multicolumn{1}{c||}{} 
  & \multicolumn{1}{c||}{}
  & \textbf{1211} & $> 4.2\,10^9$ 
  & \textbf{1211} 
  & \textbf{1211}
  & \multicolumn{1}{c|}{timeout}\\
bg 8 & 511 & 510 
  & \multicolumn{1}{c||}{}
  & \multicolumn{1}{c||}{} 
  & \textbf{32801} & $> 1.7\,10^{19}$
  & \textbf{32801}
  & \textbf{32801}
  & \multicolumn{1}{c|}{} \\
bg 9 & 1023 & 1022 
  & \multicolumn{1}{c||}{} 
  & \multicolumn{1}{c||}{}
  & \multicolumn{2}{c||}{timeout} 
  & \multicolumn{1}{c||}{timeout} 
  & \multicolumn{1}{c||}{timeout}
  & \multicolumn{1}{c|}{} \\
\hline
qo 15 & 17 & 17 
  & 6995968 
  & 3489810 
  & \textbf{407} & 32 
  & \textbf{407}
  & \textbf{407}
  & 422\\
qo 50 & 52 & 52 
  & \multicolumn{1}{c||}{timeout}
  & \multicolumn{1}{c||}{timeout}
  & \textbf{3977} & 102 
  & \textbf{3977}
  & \textbf{3977}
  & 4027\\
qo 100 & 102 & 102 
  & \multicolumn{1}{c||}{}
  & \multicolumn{1}{c||}{}
  & \textbf{15452} & 202 
  & \textbf{15452}
  & \textbf{15452}
  & \textbf{15552}\\
qo 200 & 202 & 202 
  & \multicolumn{1}{c||}{} 
  & \multicolumn{1}{c||}{}
  & \multicolumn{2}{c||}{timeout} 
  & \multicolumn{1}{c||}{timeout} 
  & \multicolumn{1}{c||}{timeout} 
  & \multicolumn{1}{c|}{timeout}\\
\hline
mpat 5 & 10 & 7 
  & 59616 
  & 11264 
  & \textbf{672} & 3840 
  & \textbf{672}
  & \textbf{672} 
  & 3202\\
mpat 10 & 20 & 12 
  & \multicolumn{1}{c||}{timeout} 
  & \multicolumn{1}{c||}{timeout} 
  & \textbf{41984} & $> 3.7\,10^9$ 
  & \textbf{41984}
  & \textbf{41984}
  & 1446066\\
mpat 15 & 30 & 17 
  & \multicolumn{1}{c||}{}
  & \multicolumn{1}{c||}{}
  & \textbf{1998848} & $> 4.2\,10^{16}$ 
  & \textbf{1998848}
  & \textbf{1998848}
  & \multicolumn{1}{c|}{timeout}\\
mpat 20 & 40 & 22 
  & \multicolumn{1}{c||}{} 
  & \multicolumn{1}{c||}{}
  & \multicolumn{2}{c||}{timeout} 
  & \multicolumn{1}{c||}{timeout} 
  & \multicolumn{1}{c||}{timeout} & \\
\hline
disp 5 & 7 & 18 
  & 95744 
  & 75786 
  & \textbf{176} & 12 
  & \textbf{176}
  & \textbf{176}
  & 224\\
disp 10 & 12 & 33 
  & \multicolumn{1}{c||}{timeout} 
  & \multicolumn{1}{c||}{timeout} 
  & \textbf{526} & 22 
  & \textbf{526}
  & \textbf{526}
  & 769\\
disp 50 & 52 & 153 
  & \multicolumn{1}{c||}{}
  & \multicolumn{1}{c||}{}
  & \textbf{9626} & 102 
  & \textbf{9626}
  & \textbf{9626}
  & 16829\\
disp 100 & 102 & 303 
  & \multicolumn{1}{c||}{} 
  & \multicolumn{1}{c||}{} 
  & \textbf{36751} & 202 
  & \textbf{36751}
  & \textbf{36751}
  & 66154\\
disp 200 & 202 & 603 
  & \multicolumn{1}{c||}{} 
  & \multicolumn{1}{c||}{} 
  & \multicolumn{2}{c||}{timeout} 
  & \multicolumn{1}{c||}{timeout} 
  & \multicolumn{1}{c||}{timeout} 
  & \multicolumn{1}{c|}{timeout}\\
\hline
tk 7 & 14 & 14 
  & 2892644 
  & 2990790 
  & \textbf{167743} & $> 1.9\,10^7$ 
  & 494259 
  & \textbf{167743}
  & \textbf{167743}\\
tk 8 & 16 & 16 
  & \multicolumn{1}{c||}{timeout}
  & \multicolumn{1}{c||}{timeout}
  & \textbf{683829} & $> 5.9\,10^8$
  & 2405450
  & \textbf{683829}
  & \textbf{683829}\\
tk 9 & 18 & 18 
  & \multicolumn{1}{c||}{} 
  & \multicolumn{1}{c||}{} 
  & \textbf{2726258} & $> 1.9\,10^{10}$ 
  & \multicolumn{1}{c||}{timeout} 
  & \textbf{2726258} 
  & \textbf{2726258}\\
tk 10 & 20 & 20 
  & \multicolumn{1}{c||}{} 
  & \multicolumn{1}{c||}{} 
  & \multicolumn{2}{c||}{timeout}
  & \multicolumn{1}{c||}{} 
  & \textbf{10702780} 
  & \multicolumn{1}{c|}{timeout}\\
tk 11 & 22 & 22 
  & \multicolumn{1}{c||}{} 
  & \multicolumn{1}{c||}{} 
  & \multicolumn{2}{c||}{} 
  & \multicolumn{1}{c||}{} 
  & \multicolumn{1}{c||}{timeout} 
  & \multicolumn{1}{c|}{} \\
\hline
pet 4 & 4 & 7 
  & 19029 
  & 23824 
  & 19584 & $> 5.7\,10^{10}$ 
  & \textbf{17648}
  & 21209 
  & 20009\\
pet 5 & 5 & 9 
  & 932573 
  & 1435970 
  & 1174715 & $> 1.6\,10^{19}$ 
  & \textbf{877567} 
  & 1295320 
  & 1201401\\
pet 6 & 6 & 11 
  & \multicolumn{1}{c||}{timeout} 
  & \multicolumn{1}{c||}{timeout} 
  & \multicolumn{2}{c||}{timeout} 
  & \multicolumn{1}{c||}{timeout}
  & \multicolumn{1}{c||}{timeout}
  & \multicolumn{1}{c|}{timeout}\\
\hline
fs 6 & 6 & 64 
  & 531441 
  & 274374 
  & 481 & 32 
  & \textbf{106}
  & 605 
  & 132 \\
fs 7 & 7 & 66 
  & \multicolumn{1}{c||}{timeout} 
  & \multicolumn{1}{c||}{timeout} 
  & 1107 & 64 
  & \textbf{177} 
  & 1359 
  & 234\\
fs 13 & 13 & 15
  & \multicolumn{1}{c||}{} 
  & \multicolumn{1}{c||}{}
  & 131103 & 4096
  & \textbf{8283}
  & 147483
  & 12366\\
fs 14 & 14 & 16
  & \multicolumn{1}{c||}{} 
  & \multicolumn{1}{c||}{}
  & 311339 & 8192 
  & 86174 
  & 364574
  & \textbf{69770}\\
fs 15 & 15 & 82 
  & \multicolumn{1}{c||}{} 
  & \multicolumn{1}{c||}{} 
  & \multicolumn{2}{c||}{timeout} 
  & \multicolumn{1}{c||}{timeout} 
  & \multicolumn{1}{c||}{timeout} 
  & \multicolumn{1}{c|}{timeout}\\
\hline
lz 12 & 12 & 14 
  & 16054352 
  & 11990059 
  & \textbf{1113570} & 190459 
  & 2030705 
  & 2654133 
  & \textbf{1113570}\\
lz 13 & 13 & 15 
  & \multicolumn{1}{c||}{timeout} 
  & \multicolumn{1}{c||}{timeout} 
  & \multicolumn{2}{c||}{timeout} 
  & \multicolumn{1}{c||}{timeout} 
  & \textbf{8319765} 
  & \multicolumn{1}{c|}{timeout}\\
lz 14 & 14 & 16 
  & \multicolumn{1}{c||}{} 
  & \multicolumn{1}{c||}{}
  & \multicolumn{2}{c||}{} 
  & \multicolumn{1}{c||}{}
  & \multicolumn{1}{c||}{timeout} 
  & \multicolumn{1}{c|}{}\\
  \hline
  dp 10 & 10 & 10
  & 9765624 
  & 5706432 
  & \textbf{145494} & $> 4.5\,10^8$ 
  & 734806 
  & 2165172 
  & 792072\\
dp 11 & 11 & 11 
  & \multicolumn{1}{c||}{timeout} 
  & \multicolumn{1}{c||}{timeout} 
  & \textbf{344511} & $> 3.5\,10^9$ 
  & 2122108 
  & 7370149 
  & 2695491\\
dp 12 & 12 & 12 
  & \multicolumn{1}{c||}{} 
  & \multicolumn{1}{c||}{} 
  & \textbf{799032} & $> 2.8\,10^{10}$ 
  & 5936405 
  & \multicolumn{1}{c||}{timeout} 
  & \multicolumn{1}{c|}{timeout}\\
dp 13 & 13 & 13 
  & \multicolumn{1}{c||}{} 
  & \multicolumn{1}{c||}{} 
  & \textbf{1822134} & $> 2.2\,10^{11}$ 
  & \multicolumn{1}{c||}{timeout} 
  & \multicolumn{1}{c||}{} 
  & \multicolumn{1}{c|}{} \\
dp 14 & 14 & 14 
  & \multicolumn{1}{c||}{} 
  & \multicolumn{1}{c||}{}
  & \textbf{4097822} & $> 1.8\,10^{12}$ 
  & \multicolumn{1}{c||}{} 
  & \multicolumn{1}{c||}{} 
  & \multicolumn{1}{c|}{} \\
dp 15 & 15 & 15 
  & \multicolumn{1}{c||}{}
  & \multicolumn{1}{c||}{}
  & \multicolumn{2}{c||}{timeout} 
  & \multicolumn{1}{c||}{} 
  & \multicolumn{1}{c||}{} 
  & \multicolumn{1}{c|}{}\\
  \hline
\end{tabular}

%% file: conclusions.tex
\section{Conclusions}

The first main contribution of this paper is a strong lower bound, indicating that stateful partial-order methods should not aim at some strict concept of optimality like
"trace-optimality" from~\cite{Abd.Aro.Jon.Sag.Source2017}.
However, we use trace-optimality as a guiding principle for the
idealized algorithm from Listing~\ref{lst:lex}.
The second contribution is the notion of $"\IFS"$ oracle and its approximations going beyond the
persistent set approach so dominant in stateful partial-order reduction
literature since its beginnings.
This $\IFS$ approach shifts the focus from deciding which actions to take to
determining when to stop. While it might seem less intuitive to ask when to stop, it is more intuitive to develop heuristics for the $\IFS$ question.

We believe that our results transfer to other models of concurrent systems.
The proof of the lower bound (Theorem~\ref{thm:por-impossible}) is intricate, but as far as we can see, it does not depend on specific features of our model.
Our new partial-order algorithm (Listing~\ref{lst:full}) utilizes the robust
and versatile framework of asynchronous automata capable of encoding various
synchronization constructs directly.
However, no formalism universally fits all concurrent systems, and certain
mismatches, like those discussed in Remark~\ref{rem:non-blocking-case} on
non-blocking, are inevitable.
Our current work is limited to binary synchronizations, but we anticipate that extending our approach to multi-party synchronizations would be feasible, albeit with more complex notation.

It is uncertain whether similar lower bounds to ours apply to stateless
partial-order reduction. Any lower bound in this scenario would have to consider algorithms operating in polynomial space. Since existing trace-optimal reductions assume some form of "non-blocking", there is no direct conflict between our lower bound and algorithms from~\cite{Kok.Mar.Gla.Vaf.Truly2022,Abd.Ati.Das.Jon.Sag.Parsimonious}.
We conjecture that an analogous lower bound might exist for a stateless
partial-order reductions in the presence of blocking.
It is also important to explore lower bounds for the unfolding approach~\cite{Rod.Sou.Sha.Kro.Unfoldingbased2015}.
Here, thanks to~\cite{Cot.Pet.Rod.Sou.Quasioptimal2021} we know a $\NP$ lower
bound for computing, so called, alternatives.
This result is similar to our result on the NP-hardness of $"\IFS"$.
Prime event structures represent runs differently than transition systems, and sometimes they can be more compact.
For this reason, our general lower bound does not
immediately translate to unfolding-based partial-order methods.

Regarding practical algorithms for stateful partial-order reductions, a crucial next step is to explore other "por-equivalence" relations.
One of the greatest forces behind modern partial-order methods comes from
their ability to work with more refined equivalences, such as allowing permutations of two reads of the
same variable and much more~\cite{Mai.Gup.Kan.Maj.Partial2016,Cha.Cha.Pav.Sin.Vai.Datacentric2018,Aga.Cha.Pat.Pav.Tom.Stateless2021,Far.Klu.Pod.Sound2022,Kok.Maj.Vaf.Enhancing2024,Far.Klu.Pod.Stratified2023,Alb.de.Gom.Isa.Stu.Optimal2023}.
We believe that our client/server formalism may offer an elegant way to
define such equivalences, and that the methodology presented here would apply.
An extension to weak memory models would also be interesting, as they have a
potential to give even bigger reductions.

One technical objective is to integrate the race reversal method.
As discussed in Remark~\ref{rem:reversals}, it can be seen as another
heuristic for $"\IFS"$, and may even be combined with the heuristic we have
proposed here.
A straightforward implementation of race reversals is linear with respect to the
number of paths in the reduced transition system.
While suitable for stateless methods, this approach is not viable in
our context, where the number of paths is often orders of magnitude bigger than the size of the
reduced system.
Developing a more efficient yet elegant algorithm is an interesting challenge.

A more general challenge is to find other heuristics for $"\IFS"$.
The $"\PIFS"$ heuristic introduced here is relatively straightforward, but it is
certainly not the only approach.
An intriguing possibility could involve
leveraging SAT solvers to find  approximations of $"\IFS"$.
Even without SAT there are many possibilities.
For instance, employing $k$-Cartesian abstractions~\cite{Hoe.Maj.Pod.Thread2017} instead of focusing solely on local
runs could offer new insights.
Additionally, in the context of verifying timed systems, methods abstracting the time component could provide novel ways to approximate $\IFS$.

Handling systems with infinite runs is another important issue, particularly for the stateful approach.
We have pointed out the source of the problem in Remark~\ref{rem:cycles}.
At present, we do not have an elegant solution for this case.